\documentclass[11pt, a4paper]{article}

\usepackage[english]{babel}
\usepackage[utf8]{inputenc}
\usepackage[authoryear]{natbib}
\usepackage{adjustbox}
\usepackage{amsmath}
\usepackage{amsthm}
\usepackage{amsfonts}
\usepackage{amssymb}
\usepackage{array}
\usepackage{graphicx}
\usepackage{floatrow}
\usepackage{caption}
\usepackage{subcaption}
\usepackage{enumitem}
\usepackage{secdot}
\usepackage{booktabs}
\usepackage{multirow}
\usepackage{multicol}
\usepackage{xr}
\usepackage[hidelinks]{hyperref}
\usepackage{bm}
\usepackage{xcolor}
\usepackage[margin=1.2in]{geometry}
\usepackage{soul}
\usepackage{tikz}
\usepackage{svg}
\usepackage{adjustbox}
\usepackage{bbm}
\usepackage{algorithm}
\usepackage[noend]{algpseudocode}
\usepackage{rotating}

\DeclareRobustCommand\solid {\tikz[baseline=-0.6ex]\draw[thick] (0,0)--(0.5,0);}

\DeclareRobustCommand\dashed{\tikz[baseline=-0.6ex]\draw[thick,dashed] (0,0)--(0.53,0);}

\newcommand{\R}{{\ensuremath{\mathbb{R}}}}

\newcommand{\cond}{\!~|~\!}

\DeclareMathOperator*{\argmax}{\arg\,\max}
\newfloatcommand{capbtabbox}{table}[][\FBwidth]

\theoremstyle{definition}

\newtheorem{condition}{Condition}[section]
\newtheorem{proposition}{Proposition}[section]

\theoremstyle{plain}
\newtheorem{example}{Example}[section]
\newtheorem{remark}{Remark}[section]

\setcitestyle{citesep={,}}

\allowdisplaybreaks

\title{\Large A Generalized Additive Partial-Mastery Cognitive Diagnosis Model}
\author{Camilo C\'ardenas-Hurtado\thanks{Department of Statistics, LSE. Contact: \href{c.a.cardenas-hurtado@lse.ac.uk}{c.a.cardenas-hurtado@lse.ac.uk}} \qquad Sze Ming Lee\thanks{Department of Statistics, LSE. Contact: \href{s.lee51@lse.ac.uk}{s.lee51@lse.ac.uk}}
\\Yunxiao Chen\thanks{Department of Statistics, LSE. Contact: \href{y.chen186@lse.ac.uk}{y.chen186@lse.ac.uk}. Corresponding author} \qquad Irini Moustaki\thanks{Department of Statistics, LSE. Contact: \href{i.moustaki@lse.ac.uk}{i.moustaki@lse.ac.uk}}}
\date{ }

\begin{document}





\maketitle
\begin{abstract}
Cognitive diagnosis models (CDMs) are restricted latent class models widely used to measure attributes of interest in diagnostic assessments across education, psychology, biomedical sciences, and related fields. Partial-mastery CDMs (PM-CDMs) are an important extension of CDMs. They model individuals' status for each attribute as continuous to measure partial mastery levels, thereby relaxing the restrictive discrete-attribute assumption of classical CDMs. As a result, PM-CDMs often yield better fits to real-world data and more refined measurements of the substantive attributes of interest. However, these models inherit strong parametric assumptions from traditional CDMs about item response functions and thus still face a significant risk of model misspecification. This paper proposes a generalized additive PM-CDM (GaPM-CDM) that substantially relaxes the parametric assumptions of PM-CDMs. This proposal leverages model parsimony and interpretability by modeling each item response function as a mixture of nonparametric monotone functions of attributes. A method for estimating GaPM-CDM is developed that combines the marginal maximum likelihood estimator with a sieve approximation of the nonparametric functions. The new model is applicable in both confirmatory and exploratory settings, depending on whether prior knowledge of the relationship between observed variables and attributes is available. The proposed method is evaluated and compared with PM-CDMs through extensive simulation studies and further
applied to two measurement problems from educational testing and healthcare research, respectively.

\medskip

\end{abstract}

\newpage

\section{Introduction} \label{sec:Introduction}

Diagnostic assessments are commonly used in education, psychology, biomedical science, and related fields to identify individuals' attribute profiles based on their observed responses to assessment items. An attribute refers to an individual's latent dimension, such as a problem-solving skill, a knowledge component, a personality trait, or a mental health disorder. An attribute profile typically involves multiple attributes that are likely correlated. Measuring attribute profiles is a non-trivial task due to the potentially complex relationships between attributes and items, as well as the dependence between attributes.

Cognitive diagnostic models (CDMs), with roots in the rule-space model \citep{Tatsuoka_JEM1983} and latent class analysis \citep{Book_LH1986}, have been proposed to tackle the measurement challenge with diagnostic assessments. Various CDMs have been developed under different assumptions about the cognitive process or characteristics of the latent attributes (e.g., \citealp{Junker&Sijtsma_APM2001, vonDavier_BJMSP2008, HensonEtAl_Psychometrika2009, delaTorre_Psychometrika2011, Chen&deLaTorre_APM2013, ZhanEtAl_JoClass2020, Ma_MBR2022}); see \cite{Book_RuppEtAl2010} and \cite{Book_vDL2019} for a comprehensive review. The attributes in the traditional CDMs are assumed to be discrete. Early developments of CDMs typically consider binary attributes, under which an individual either fully masters or does not master an attribute. In subsequent developments of CDMs, the binary-attribute assumption has been relaxed in several models. These models allow each attribute to have a small number of ordered levels, which, however, still may not be sufficient to support fine-grained inference on individuals' partial mastery levels on attributes. To fill this gap, \cite{ShangEtAl_AoAS2021} proposed a flexible family of partial-mastery CDMs (PM-CDMs), which assumes the attributes to be continuous rather than discrete. More specifically, an individual is assumed to have a set of continuous partial mastery scores between 0 and 1 that measure their mastery level for each attribute, where a larger score represents a higher level of partial mastery, and the extreme scores of zero and one represent complete non-mastery and mastery, respectively. It was found in \cite{ShangEtAl_AoAS2021} that the use of continuous partial-mastery scores tends to yield better fits for real-world data than classical CDMs and refined measurement of the substantive attributes of interest. Technically, the PM-CDMs are developed by combining parametric assumptions of traditional CDMs with modeling techniques from the grade-of-membership model \citep{Erosheva2002}, also known as the mixed-membership model \citep{Book_MMM2014}, a general family of models that have been widely used to model complex multivariate data in computer sciences, social surveys, genetics, among other fields (see, e.g., \citealp{BleiEtAl_JMLR2003, EroshevaEtAl_AoAS2007, AiroldiEtAl_JMLR2008}). By assuming the mastery level to be continuous, PM-CDMs also establish a link between CDMs and multidimensional item response theory (IRT) models \citep[see, e.g.,][for a review]{ChenEtAl_StatSci2025}, a general family of latent variable models tailored for analyzing item response data. 

Although the PM-CDMs have demonstrated significant advantages, they remain constrained by inheriting relatively strong parametric assumptions from classical CDMs about the item response functions (IRFs), i.e., the conditional distributions of the manifest variables given the partial mastery scores. To mitigate this restriction, \cite{ShangEtAl_AoAS2021} gave multiple options for the parametric IRFs. However, selecting the IRF is a model selection problem that can be challenging for practitioners, especially when different items are allowed to have different parametric forms for the IRF. On the other hand, it has long been observed in the IRT literature that conventional parametric families sometimes fail to capture patterns in many real-world item response data, and nonparametric and semiparametric IRT models have been proposed as a remedy \citep{Ramsay_StatSci1988, Ramsay&Winsberg_Psychometrika1991, Book_SM2002}. However, all the existing nonparametric and semiparametric IRT models are unidimensional, in the sense that only one-dimensional latent trait/attribute is modeled. Extending them to the multidimensional setting of PM-CDMs is the focus of this paper. Such an extension is not straightforward, as we discuss in the sequel.

This paper proposes a Generalized Additive Partial-Mastery CDM (GaPM-CDM) as a semiparametric PM-CDM. Instead of a fully nonparametric multivariate function, the new model is kept parsimonious by assuming each IRF to have a semiparametric generalized additive form, similar to the generalized additive model used in regression analysis \citep{Book_HT1990}. More specifically, each IRF is assumed to be a mixture of nonparametric monotone functions of the partial mastery scores of attributes with nonnegative mixture weights. Both the monotonicity assumption and the nonnegative weights are important for interpreting the IRFs. The former captures the monotone relationship between an attribute and an item, conditioning on the rest of the attributes. Similar monotone assumptions are imposed in parametric CDMs \citep{HensonEtAl_Psychometrika2009, FangEtAl_Psychometrika2019, ShangEtAl_AoAS2021} as well as IRT models \citep[e.g.,][]{Ramsay_StatSci1988, Ramsay&Winsberg_Psychometrika1991}. The nonnegative weights of an item capture the contributions of the attributes to the response, generalizing the concept of \textbf{Q}-matrix in classical CDMs. Specifically, a zero weight indicates the conditional independence between the corresponding attribute and item response, given the remaining attributes. The use of the nonnegative weights also allows us to apply the GaPM-CDM in an exploratory setting when the item-attribute relationship, i.e., the \textbf{Q}-matrix, is unknown, while \cite{ShangEtAl_AoAS2021} only considered a confirmatory setting with a known \textbf{Q}-matrix. The proposed model is also a multidimensional extension of the semiparametric IRT model proposed in \cite{Ramsay&Winsberg_Psychometrika1991}. 

To estimate the GaPM-CDM, we propose a sieve marginal maximum likelihood estimator that combines the `method of sieves' \citep{Shen_AoS1997} for infinite-dimensional nonparametric functions with the standard marginal maximum likelihood estimator for latent variable models \citep{Book_SR2004, Book_BKM2011}. In particular, we use piecewise linear functions to approximate the monotone IRFs. To handle the computational challenge posed by the integrals with respect to the latent variables (i.e., attributes) in the marginal likelihood, a computationally efficient Markov chain Monte Carlo stochastic-approximation (MCMC-SA) algorithm \citep{Robbins&Monro_AoMS1951, Gu&Kong_PNAS1998, Zhang&Chen_Psychometrika2022} is developed. This algorithm iteratively constructs stochastic gradients of the marginal log-likelihood by sampling the latent variables with a Markov chain Monte Carlo (MCMC) sampler, and then updates the unknown parameters using these gradients. The proposed model is validated via extensive simulation studies and further applied to two real-world datasets.
 
The rest of the paper is organized as follows. Section \ref{sec:Data} introduces two case studies, motivating an extension of PM-CDMs robust to misspecifications in both the parametric forms of the IRFs and the \textbf{Q}-matrix. Section~\ref{sec:GaPM-CDM} gives a brief review of PM-CDMs and then introduces the GaPM-CDM framework and discusses the estimation of the model parameters. Section~\ref{sec:Sims} presents extensive simulation studies comparing the performance of GaPM-CDM against comparable PM-CDMs and CDMs in terms of IRF recovery and latent attribute scoring. Section~\ref{sec:App} presents analyses of the English test and patient-reported outcomes data sets. Section~\ref{sec:Disc} concludes and provides further discussion. We also provide Online Supplementary Materials, including technical details of the estimation strategy, simulation studies, additional results on the empirical applications, and theoretical results on model identifiability.

\section{Motivating case studies} \label{sec:Data}

We present two real-world applications that motivate the development of the GaPM-CDM. The first, drawn from educational testing, concerns a classic benchmark dataset with a predefined $\mathbf{Q}$-matrix. Despite its extensive use in the literature, our analysis uncovers new insights regarding nonlinear IRFs and potential $\mathbf{Q}$-matrix misspecification. The second application analyzes patient-reported outcomes from clinical healthcare research to assess social-role performance. In this case, both the number of attributes and the item-attribute relationships are unknown and are learned directly from the data.

\subsection{English Test Data} \label{sec:Data_ECPE}

The English test dataset comes from the 2003-2004 grammar section of the Examination for the Certificate of Proficiency in English (ECPE), designed and administered by the University of Michigan English Language Institute. The dataset contains responses to items evaluating latent attributes related to domains of English grammar. A \textbf{Q}-matrix is available based on the test design. This dataset has been previously analyzed in \citet{Templin&Bradshaw_Psychometrika2014, ChiuEtAl_Psychometrika2016} and \citet{ShangEtAl_AoAS2021} using classical CDMs and PM-CDMs. Results in Section \ref{sec:App_ECPE} show that the proposed model fits this data better than traditional PM-CDMs and produces flexible and interpretable IRFs that capture guessing and slipping behaviors common in educational testing, without affecting the ranking of test takers on their latent abilities. Moreover, the estimated weight parameters suggest potential \textbf{Q}-matrix misspecification for some items, further demonstrating the usefulness of the proposed method.

\subsection{Patient-reported Outcomes Data} \label{sec:Data_PROMIS}

The Patient-Reported Outcomes Measurement Information System (PROMIS; \citealp{CellaEtAl_JoCE2010}) is an interdisciplinary university consortium initiative advocating for standardized, precise, and efficient measurement of patient-reported symptoms. The PROMIS network collects self-reported data on physical, mental, and social health from a representative sample of the general United States population and multiple clinical populations. We study the module on social health, with an emphasis on social-role performance, a hypothesized multidimensional construct that measures an individual's ability to engage in and participate in daily life activities \citep{CastelEtAl_QoLR2008, HahnEtAl_SIR2010}. In this case, both the item-attribute relationship (i.e., the \textbf{Q}-matrix) and the number of attributes are unknown. Section \ref{sec:App_PROMIS} shows that the GaPM-CDM recovers a reasonable number of interpretable attributes related to social, work, and family functioning, with improved fit over traditional PM-CDMs.

\section{A Generalized Additive Partial-Mastery CDM} \label{sec:GaPM-CDM}
 
\subsection{Review of PM-CDM} \label{sec:PM-CDM} 

Following \cite{ShangEtAl_AoAS2021}, consider a diagnostic setting where individuals respond to a set of $J$ binary items, denoted by $\mathbf{Y} =(Y_j: j = 1,\ldots, J)^\top \in \{0,1\}^J$. The PM-CDM assumes that the distribution of the responses is determined by a vector of continuous latent variables $\mathbf{U} = (U_k: k = 1,\ldots,K)^\top\in [0,1]^K$ that indicate the partial mastery levels for $K$ attributes of interest, where $U_k = 0$ and $1$ denote the lowest and highest mastery levels for attribute $k$, respectively. It further assumes a known matrix $\mathbf{Q} = (q_{jk})_{J\times K}$ that characterizes the item-attribute relationship, where $q_{jk} = 1$ indicates that the $j$-th item directly measures the $k$-th attribute and $q_{jk} = 0$ otherwise. We further use $\bm q_j = (q_{j1}, \ldots, q_{jK})^\top$ to denote the $j$-th row of $\mathbf Q$. 

A PM-CDM is a parametric model with the following structure. The observed responses satisfy the \textit{local independence} assumption, which states that item responses are independent conditional on the latent variables. Moreover, responses are distributed $Y_{j} \cond \mathbf{U} \sim \mathrm{Bernoulli}(\pi_{j}(\mathbf{U}))$, where the probability $\pi_{j}(\mathbf{U}) := \mathbb{P}(Y_{j} = 1 \cond \mathbf{U})$ is a parametric function of the latent variable scores $\mathbf{U}$ that inherits assumptions of classical CDMs. For example, Example \ref{ex:aPM-CDM} below gives the additive Partial-Mastery CDM (aPM-CDM), also referred to as the partial-mastery additive CDM in \cite{ShangEtAl_AoAS2021}, whose IRF follows from the additive CDM (ACDM; \citealp{delaTorre_Psychometrika2011}). This model is closely related to the GaPM-CDM, which will be introduced in the sequel. Other parametric forms of $\pi_j(\mathbf{U})$ are given in \cite{ShangEtAl_AoAS2021} based on other classical CDMs, such as the Deterministic Input Noisy output “And” gate model (DINA; \citealp{Junker&Sijtsma_APM2001}), the Deterministic Input Noisy output “Or” gate model (DINO; \citealp{Templin&Henson_PsychMeth2006}), and the generalized DINA model (G-DINA; \citealp{delaTorre_Psychometrika2011}). 

\begin{example}\label{ex:aPM-CDM}
The IRF of the aPM-CDM takes a linear form: 
\begin{equation}\label{eq:aPM-CDM}
    \pi_{j}(\mathbf{U}) = \delta_{j0} + \sum\limits_{k=1}^K \delta_{jk} q_{jk} U_k, \qquad j =1, \ldots, J,
\end{equation}
where $\bm\delta_j = (\delta_{j0},\delta_{j1},\ldots,\delta_{jK})^\top$ are item-specific parameters. The IRF $\pi_j(\mathbf{U})$ in \eqref{eq:aPM-CDM} depends on the latent attribute $k$ when $q_{jk} = 1$. For these dimensions, the parameters $\delta_{jk}$ are typically assumed to be nonnegative to impose a monotone relationship between the partial mastery scores and the item response probabilities. Consequently, it is guaranteed that $\pi_{j}(\mathbf{U}) \geq \pi_{j}(\tilde{\mathbf{U}})$, when $U_k \geq \tilde U_k$ for all the relevant dimensions $k$, a monotone constraint commonly assumed in PM-CDMs. Moreover, as $\pi_j(\mathbf{U})$ takes a value between 0 and 1, the parameters are naturally subject to the constraints $\delta_{j0} \geq 0$ and $ \delta_{j0} + \sum_{k=1}^K \delta_{jk} q_{jk} \leq 1$.
\end{example}

The latent mastery scores are assumed to follow a Gaussian copula model. That is, each $U_k \sim \mathrm{Uniform}(0,1)$, and their joint cumulative distribution function is:
\begin{equation*}
    D(\mathbf{U}; \bm{\mu}, \bm{\Sigma}) = \bm{\Phi}(\Phi^{-1}(U_1), \ldots, \Phi^{-1}(U_K); \bm{\mu}, \bm{\Sigma})\,,
\end{equation*}
where $\Phi^{-1}$ denotes the inverse of the standard Normal cumulative distribution function, and $\bm{\Phi}(\cdot; \bm{\mu}, \bm{\Sigma})$ is the multivariate Normal cumulative distribution function with mean vector $\bm{\mu} \in \R^K$ and positive semi-definite covariance matrix $\bm{\Sigma} = (\sigma_{kk'})_{K \times K}$. The specifications above yield a joint distribution over the item responses $\mathbf{Y}$ and the latent variables $\mathbf{U}$, from which we can derive the marginal likelihood function and estimate the model parameters. 

\subsection{Model Specification of GaPM-CDM}\label{subsec:proposed}

We now propose a GaPM-CDM, which has the same assumptions as the PM-CDMs reviewed previously, except that its IRF is assumed to take a generalized additive form as in \cite{Book_HT1990}. That is, 
\begin{equation}
 \pi_{j}(\mathbf{U}) = \sum\limits_{k = 1}^{K} \alpha_{jk}q_{jk} g_{jk}(U_{k})\,, \qquad j = 1, \ldots, J, \label{eq:P(Y_j|U) GaPM-CDM}
\end{equation}
where $g_{jk}: [0,1] \rightarrow [0,1]$ are continuous, monotone non-decreasing functions, and $\alpha_{jk}$s are the associated nonnegative weights. To ensure $\alpha_{jk}$ and $g_{jk}$ are identifiable, we impose the boundary conditions $g_{jk}(0) = 0$ and $g_{jk}(1) = 1$, and the constraint $\sum_{k = 1}^{K} \alpha_{jk}q_{jk} = 1$. Moreover, we set $\bm{\mu}$ to be a zero vector and the diagonal entries of $\bm{\Sigma}$ to take values of one in the Gaussian copula model, as they cannot be identified due to the flexibility of the monotone functions $g_{jk}$. As a result, each partial-mastery score $U_k$ now marginally follows a uniform distribution on the interval $[0,1]$. We shall note that the off-diagonal entries of $\bm{\Sigma}$ are still estimated to learn the dependence between the attributes.

\begin{remark}[Comments on the IRF of GaPM-CDM]
First, the IRF in \eqref{eq:P(Y_j|U) GaPM-CDM} satisfies the same monotone constraint as the aPM-CDM and many other PM-CDMs in that $\pi_{j}(\mathbf{U}) \geq \pi_{j}(\tilde {\mathbf{U}})$, when $U_k \geq \tilde U_k$ for all $k$ such that $q_{jk} = 1$. Second, this IRF also enforces some boundary conditions. That is, $\pi_j(\mathbf{U}) = 0$ when $U_k = 0$ for all $k$ such that $q_{jk} = 1$, and $\pi_j(\mathbf{U}) = 1$ when $U_k = 1$ for all $k$ such that $q_{jk} = 1$. That means, when an individual does not master any of the attributes required to solve an item, they have zero chance of correctly answering it. On the other hand, when an individual fully masters all relevant attributes, it is 100\% certain that they can correctly answer the item. Compared to the aPM-CDM and other PM-CDMs, our IRF does not model the guessing and slipping probabilities, which are the probabilities of correctly answering the item when an individual does not master any relevant attributes and incorrectly answering it when the individual fully masters all the relevant attributes, respectively. However, the nonparametric functions $g_{jk}$ can approximate near-guessing and near-slipping behavior arbitrarily close to the boundaries of the latent attribute space (further details in the Online Supplementary Materials). Third, we note that many PM-CDMs, such as the partial-mastery DINA model, model interactions among attributes, which the IRF of the GaPM-CDM cannot fully capture. We believe it is possible to incorporate interaction terms in a semiparametric fashion into \eqref{eq:P(Y_j|U) GaPM-CDM}. As the current model is already complex, we leave this extension to future research; see Section~\ref{sec:Disc} for further discussion.
\end{remark}

\begin{remark}[Confirmatory versus exploratory settings]
When introducing the IRF for the GaPM-CDM, we follow the confirmatory setting as in \cite{ShangEtAl_AoAS2021}. However, the proposed model can also be applied under exploratory data analysis settings when the $\mathbf{Q}$-matrix is unknown. In that case, the IRF takes the form
\begin{equation}
 \pi_{j}(\mathbf{U}) = \sum\limits_{k = 1}^{K} \alpha_{jk} g_{jk}(U_{k})\,, \qquad j = 1, \ldots, J, \label{eq:GaPM-CDMexp}
\end{equation}
which is equivalent to setting $q_{jk} = 1$ for all $j$ and $k$ in \eqref{eq:P(Y_j|U) GaPM-CDM}. Unlike certain multidimensional IRT models and CDMs that have indeterminacy issues when the $\mathbf{Q}$-matrix imposes no constraint \citep[see, e.g.,][]{ChenEtAl_JASA2020, Gu&Xu_StatSinica2021}, the GaPM-CDM does not seem to suffer from similar indeterminacy under the exploratory setting, as discussed in Remark \ref{rmk:identif} and empirically verified in the simulation study in Section \ref{sec:Sims_S2}.
\end{remark}

\begin{remark}[Connection to semiparametric IRT models]
The proposed GaPM-CDM is an extension of the semiparametric IRT model in \citet{Ramsay&Winsberg_Psychometrika1991} to a multidimensional setting. In fact, it can be shown that the model in \citet{Ramsay&Winsberg_Psychometrika1991} is mathematically equivalent to the proposed GaPM-CDM when the latent dimension $K=1$ and $q_{j1} = 1$ for all $j = 1, \ldots, J$. 
\end{remark}

\subsection{Estimation} \label{sec:Estimation}
In what follows, we consider the estimation of GaPM-CDM, given observed data from $N$ individuals, denoted by $\mathbf{y}_i= (y_{ij}:j = 1, \ldots, J)^\top$, $i = 1,\ldots, N$. To handle the infinite-dimensional functions $g_{jk}$ in the model, we propose an estimator that combines the marginal maximum likelihood estimator with a sieve approximation of the infinite-dimensional functions, in which each $g_{jk}$ admits an approximation $g_{jk}^s$ in a finite-dimensional sub-space $\mathcal{G}_L$. The approximation error decreases as the sieve $\mathcal{G}_L \subset \mathcal{G}_{L+1} \subset \cdots$ becomes dense in the original infinite-dimensional parameter space \citep{Book_Grenander1981, Shen_AoS1997}. For simplicity and interpretability, we model $g_{jk}^s$ through piecewise linear functions, although other bases for monotone non-decreasing continuous functions can be accommodated as well (e.g., I-Splines \citealp{Ramsay_StatSci1988}).

\begin{figure}[!ht]
\centering
\vspace{5mm}
\includegraphics[scale = 0.6]{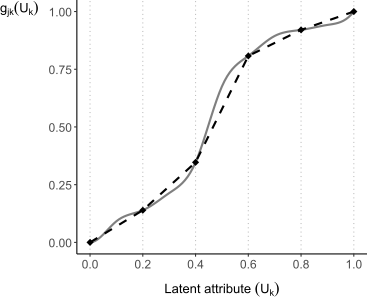}
\caption{Example of a continuous monotone function $g_{jk}$ (solid line, \solid) and its sieve piecewise linear approximation $g_{jk}^s$ (dashed line, \dashed), evaluated on equally spaced inner knots $\bm{\kappa} = (0.2,0.4,0.6,0.8)^\top$.}
\label{fig:IRF_example}
\end{figure}

More specifically, let $\bm{\kappa} = (\kappa_{l}: l = 1,\ldots, L)^\top$ be a vector of fixed inner grid points such that $0 = \kappa_0 < \kappa_{1} < \cdots < \kappa_{L} < \kappa_{L+1} = 1$, with $\{\kappa_0,\,\kappa_{L+1}\}$ fixed as boundary knots. The piecewise linear approximation of $g_{jk}$ is given by:
\begin{equation}
g_{jk}^s(x; \bm{\theta}_{jk},\bm{\kappa}) = \begin{cases}
    \dfrac{\theta_{jk,1}}{\kappa_{1}}\, x & \mathrm{if} \quad x \in [0,\kappa_1), \\[3pt]
    \theta_{jk,1} + \dfrac{\theta_{jk,2}}{\kappa_2 - \kappa_{1}}\, (x - \kappa_{1}) & \mathrm{if} \quad x \in [\kappa_1,\kappa_2), \\[3pt]
    & \vdots \\
    \left(\sum_{l=1}^{L} \theta_{jk,l}\right) + \dfrac{\theta_{jk,L+1}}{1 - \kappa_{L}}\, (x - \kappa_{L}) & \mathrm{if} \quad x \in [\kappa_{L},1],
    \end{cases} \label{eq:g_jk(N)}
\end{equation}
where $\bm{\theta}_{jk} = (\theta_{jk,l}: l = 1,\ldots,L+1)^\top$ is a vector of unknown approximation parameters satisfying the constraints $\theta_{jk,l} \geq 0$ for all $l = 1,\ldots,L+1$ and $\sum_{l=1}^{L+1} \theta_{jk,l} = 1$. Figure \ref{fig:IRF_example} shows an example of a monotone function $g_{jk}$ and its piecewise linear approximation $g_{jk}^s$. 

\begin{remark}[Identifiability] \label{rmk:identif}
Generic identifiability \citep{AllmanEtAl_AoS2009, Gu&Xu_AoS2020}, where model parameters are identifiable for almost all points in the parameter space except for a subset of Lebesgue measure zero, can be established for the GaPM-CDM with $g_{jk}^s$ defined in the class of non-decreasing step functions, $g^{\text{step}}_{jk}$. This class approximates any $g_{jk}^s$-- and thus the original $g_{jk}$-- arbitrarily well as the grid becomes finer (further details in Online Supplementary Material). We require the following condition:
\begin{condition}[\textbf{Q}-matrix structure] \label{cond:condQ}
    The \textbf{Q}-matrix is of the form $\mathbf{Q}^{\top} = (\mathbf{Q}_1^\top, \mathbf{Q}_2^\top,(\mathbf{Q}')^{\top})$, where, for $i \in \{1,2\}$, $\mathbf{Q}_i = (q_{i,jk})_{(KL) \times K}$ has entries $q_{i,jk} = 1$ in positions $(j,k)$ such that $j = K(c-1)+k$, with $c=1,\dots,L$, and $k = 1, \dots, K$; and zero or one elsewhere. In other words, sub-matrices $\mathbf{Q}_i$, $i \in \{1,2\}$, are constructed by vertically stacking $K \times K$ matrices with diagonal entries being one, $L$ times. Each column of the sub-matrix $\mathbf{Q}'$ has at least one non-zero entry.
\end{condition}

\begin{proposition}[\textit{Informal}] \label{prop:identif}
In confirmatory settings, if the \textbf{Q}-matrix satisfies Condition~\ref{cond:condQ}, then the parameters characterizing a sieve approximation of the IRFs in the GaPM-CDM are generically identified. In exploratory settings, a sufficient condition for generic identifiability is $J \geq 2KL + 1$.
\end{proposition}

Our proof of Proposition \ref{prop:identif} leverages a restricted latent class model (RLCM) representation of the GaPM-CDM with IRFs approximated by piecewise constant functions (which, in turn, can approximate monotone non-decreasing continuous functions arbitrarily well). In particular, we show that parameters characterizing such IRFs can be mapped to a corresponding set of RLCM parameters for which generic identifiability holds up to label swapping \citep{Gu&Xu_AoS2020}. A formal statement for Proposition \ref{prop:identif}, definition of generic identifiability, and detailed proofs are provided in the Online Supplementary Material. While Proposition \ref{prop:identif} does not directly address generic identifiability of $\bm{\Sigma}$, we show that the induced distribution of the latent classes under $\bm{\Sigma}$ in the RLCM representation is generically identifiable (see Proposition S1 in the Online Supplementary Materials).
\end{remark}

An approximate marginal log-likelihood can be written as 
\begin{equation}
    \ell(\bm{\Theta}_{\bm\kappa}) = \sum\limits_{i = 1}^N \log \left(~ \int\limits_{[0,1]^K} \prod\limits_{j = 1}^J \pi_{j}(\mathbf{U}; \bm{\alpha}_j, \bm{\theta}_{j}, \bm{\kappa})^{y_{ij}} \, (1-\pi_{j}(\mathbf{U}; \bm{\alpha}_j, \bm{\theta}_{j}, \bm{\kappa}))^{1-y_{ij}}\, \mathrm{d}D(\mathbf{U}; \bm{\Sigma}) \right), \label{eq:mllk(U)}
\end{equation}
where $\bm\alpha_j =(\alpha_{jk}:k=1,\ldots,K)^\top$ and $\bm\theta_j = (\bm\theta_{jk}:k = 1,\ldots,K)^\top$ are the weights and parameters in the sieve approximation of the IRF for item $j$ given by $\pi_{j}(\mathbf{U}; \bm{\alpha}_j, \bm{\theta}_{j}, \bm{\kappa}) = \sum_{k = 1}^{K} \alpha_{jk}q_{jk} g_{jk}^s(U_k; \bm{\theta}_{jk}, \bm{\kappa})$, $D(\mathbf{U}; \bm{\Sigma}) = D(\mathbf{U}; \bm{0}, \bm{\Sigma})$ denotes the cumulative distribution function for the Gaussian copula model, for which the mean and covariance matrix satisfy the constraints introduced in Section \ref{subsec:proposed}, and $\bm{\Theta}_{\bm{\kappa}}$ is introduced as a generic notation for the vector of unknown parameters. For computational convenience, we parameterize $\bm{\Sigma}$ through its Cholesky decomposition $\bm{\Sigma} = \mathbf{L}\mathbf{L}^\top$, where $\mathbf{L} = (l_{kk'})_{K \times K}$ is a lower triangular matrix with rows denoted by $\bm{l}_k^\top$. Thus, $\bm\Theta_{\bm\kappa} = (\bm\alpha_{j}, \bm{\theta}_{jk}, \bm l_{k}:j = 1,\ldots, J;\, k = 1,\ldots,K)^\top$. We include $\bm\kappa$ as a subscript to emphasize its dependence on the grid points.

Note that all the unknown parameters are subject to constraints. In particular, $\alpha_{jk}$ and $\theta_{jk,l}$ are nonnegative and $\bm\alpha_{j}^\top \bm q_{j} = 1$ and $\bm \theta_{jk}^\top \bm 1 = 1$, where $\bm 1$ is a vector of appropriate dimension with all elements being 1. In addition, rows of $\mathbf L$ must satisfy $||\bm l_{k}||^2_2 = 1$, where $||\cdot||_2$ denotes the Euclidean norm, to ensure $\bm{\Sigma}$ is a positive definite correlation matrix. We denote the constrained parameter space of $\bm\Theta_{\bm\kappa}$ by $\bm{\Xi}_{\bm \kappa}$. We estimate $\bm\Theta_{\bm \kappa}$ by the sieve marginal maximum likelihood estimator (SMMLE) that maximizes the approximate marginal log-likelihood, 
\begin{equation}
\hat{\bm\Theta}_{\bm \kappa} = \argmax_{\bm{\Theta}_{\bm \kappa}\, \in\, \bm{\Xi}_{\bm \kappa}} ~\ell(\bm\Theta_{\bm \kappa}). \label{eq:ThetaHat}
\end{equation}

\begin{remark}[Comment on the asymptotic consistency of SMMLE] \label{rmk:estimator}
We note that the standard asymptotic theory for sieve estimators \citep{Shen_AoS1997} is not directly applicable to the SMMLE \eqref{eq:ThetaHat} under the conventional asymptotic regime where the number of items $J$ is fixed and the sample size $N$ goes to infinity. This is because the data we consider here are binary. As the individuals are assumed to be independent and identically distributed in our model, the maximum number of parameters that can be estimated is $2^J-1$. When $J$ is fixed, it is impossible to estimate the infinite-dimensional functions consistently. As a large sample size is also required to estimate nonparametric functions, this implies that the proposed model is more suitable for large-scale data when both the number of items, $J$, and the sample size, $N$, are large. In fact, we believe that it is possible to establish the consistency of the SMMLE under a double asymptotic regime where both $N$ and $J$ grow to infinity, a setting commonly considered in the analysis of nonparametric item response theory models \citep{Douglas_Psychometrika1997} and high-dimensional latent variable models \citep{ChenEtAl_JASA2020, Chen&Gu_Psychometrika2024}. A plausible asymptotic regime requires $N,J \to \infty$, and $L \to \infty$ at a suitable speed relative to both $N$ and $J$ to balance the bias-variance tradeoff in the estimation of the IRFs. Indeed, from Remark \ref{rmk:identif}, $L$ growing at the same order of $J$ is sufficient for generic identifiability, and thus, under additional regularity conditions, for the consistency of the SMMLE. We leave a formal derivation of the consistency result for future investigation.
\end{remark}

\begin{remark}[Number of knots and their placement] \label{rmk:knots}
    Selecting the placement and number of knots ($L$) is a complex combinatorial task that involves a trade-off between approximation quality and model complexity, and simplifications are therefore necessary in practice. In this paper, we fix both $L$ and the knot positions using equally spaced, boundary-focused grids. These choices serve different purposes and are guided by prior knowledge of the expected shapes of the IRFs from the simulation studies in Section \ref{sec:Sims} and the empirical applications in Section \ref{sec:App}. In practice, researchers can choose $L$ and the placement of knots following basic guidelines. Without a penalty term in the marginal log-likelihood, $L$ becomes a smoothing parameter to be tuned by addressing the bias-variance trade-off. From Remark~\ref{rmk:identif}, $L = (J-1)/2K$ is the lower bound for the number of knots that preserves identifiability of the GaPM-CDM; however, any $L$ that grows no faster than the rate suggested in Remark~\ref{rmk:estimator} is valid. Regarding knot placement, alternative grids can be constructed through a bijective transformation $m:(0,1) \to (0,1)$, e.g., the Beta($\alpha,\beta$) cumulative distribution function with both $\alpha$ and $\beta$ larger than 1, whose output yields knots concentrated near the endpoints of the interval. This choice might produce more stable IRFs when the posterior distribution of latent attributes is concentrated near the boundaries. More generally, practitioners may consider cross-validated adaptive or data-driven strategies for knot placement.
\end{remark}

\subsection{Computation} \label{sec:Computation}

Solving the maximization problem \eqref{eq:ThetaHat} is computationally nontrivial. There are two challenges. First, the approximate marginal log-likelihood involves a $K$-dimensional integral in the $K$-dimensional cube. The computational complexity of a standard expectation-maximization (EM) algorithm \citep{DempsterEtAl_JRSS1977} grows exponentially fast with the latent dimension, which quickly becomes computationally infeasible when $K \geq 5$. Second, the unknown parameters live in a constrained parameter space $\bm{\Xi}_{\bm \kappa}$. Therefore, care must be taken to ensure that the updated parameters remain in $\bm{\Xi}_{\bm \kappa}$ at all times. To tackle both issues, we propose a stochastic-approximation mirror descent (SA-MD) algorithm that combines the stochastic-approximation algorithms for latent variable models \citep{DeBortoliEtAl_Stats&Comp2021, Zhang&Chen_Psychometrika2022} with mirror gradient descent (\citealp{Book_NY1983, Beck&Teboulle_ORL2003}). This algorithm iterates between two steps -- 1) a stochastic-approximation (SA) step that constructs an approximate stochastic gradient of $\ell(\bm\Theta_{\bm \kappa})$ by generating approximate samples of the latent variables from their posterior distributions, and 2) a mirror-descent (MD) step that updates $\bm\Theta_{\bm\kappa}$ using the approximate stochastic gradient from the SA step within the constrained space $\bm\Xi_{\bm\kappa}$. The proposed algorithm scales linearly in $N$, $J$, and $L$, and quadratically in $K$, as opposed to quadrature-based EM algorithms that depend exponentially on $K$. Due to space constraints, we reserve technical and implementation details and computational complexity analysis of the SA-MD algorithm to the Online Supplementary Materials. The proposed method has been implemented in the \texttt{R} package \texttt{gapmCDM}, available online at \href{https://github.com/ccardehu/pmCDM}{https://github.com/ccardehu/gapmCDM}.

\begin{remark}[Factor scores] \label{rmk:fscores}
The latent attributes for individual $i$ are sampled from the approximate posterior distribution $\mathbf{U}_i^{(t)} \sim f(\mathbf{U}\cond \mathbf{y}_i; \bm\Theta^{(t)}_{\bm\kappa})$ at iterations $t = 1,\ldots, T$ of the proposed SA-MD algorithm. Thus, the expected a-posteriori (EAP) factor scores can be computed as the Polyak-Ruppert average $\hat{\mathbf{U}}_i = \frac{1}{T-\omega} \sum_{t = \omega+1}^T \mathbf{U}^{(t)}_i$, for all $i = 1,\ldots,N$, where $\omega < T$ is a fixed burn-in period.
\end{remark}

\begin{remark}[Model comparison] \label{rmk:comparison}
Standard methods for model comparison cannot be applied to the proposed model. First, true parameters in the GaPM-CDM can live in the boundaries (e.g., $\alpha_{jk}$ can take values of $0$ or $1$), which violates standard conditions of the likelihood ratio test (see also \citealp{ChenEtAl_Psychometrika2020b}). Second, information criteria are not well-defined in the context of infinite-dimensional latent variable models. While the sieve approximation uses $JK(L+2) + K(K-1)/2$ parameters, the effective dimensionality (degrees of freedom) is lower due to the IRF shape and parameter constraints. Designing an appropriate information criterion with a penalty function that accounts for the dimensionality of the sieve approximations and their bias is nontrivial. For model comparison, we use cross-validation (CV) techniques with a focus on predictive performance on out-of-sample data.
\end{remark}

\section{Simulation Studies} \label{sec:Sims}
In what follows, we evaluate the proposed model via two simulation studies. The first study concerns a confirmatory setting, for which the $\mathbf{Q}$-matrix and the number of latent attributes are known. The proposed model is evaluated in terms of model fitting and parameter estimation, and further compared with the aPM-CDM. The second study considers an exploratory setting where the $\mathbf{Q}$-matrix is unknown. In this setting, we examine whether the attributes can be accurately measured without knowledge of the $\mathbf{Q}$-matrix. 

\subsection{Study I: Confirmatory Setting} \label{sec:Sims_S1}

\subsubsection{Data Generation}

We present two simulation settings, one in which the data is generated from an aPM-CDM, and one from a GaPM-CDM. In both cases, we consider a fixed-length test with $J = 20$ items, two sample sizes $N \in \{1000,3000\}$, and number of latent attributes $K \in \{3,5\}$. To make models comparable, latent attributes are generated from a Gaussian copula with mean $\bm{\mu} = \bm{0}_K$ and correlation matrix $\bm{\Sigma} = \sigma\bm{1}_K\bm{1}_K^\top + (1-\sigma)\mathbb{I}_K$, where $\mathbb{I}_K$ is the $K$-dimensional identity matrix, for $\sigma \in \{0,0.7\}$. Thus, in total, we consider 16 simulation scenarios, 8 per model. In each case, we generate $R = 100$ datasets. The \textbf{Q}-matrices used in this study are included in the Online Supplementary Materials.

When the data comes from the aPM-CDM, we set the guessing and slipping probabilities to at most $0.2$ on all IRFs. More specifically, we generate intercepts from $\delta_{j0} \sim \mathrm{Uniform}(0,0.2)$ and random slopes such that $1 - \sum_{k=1}^K \delta_{jk} q_{jk} \leq 0.2$ for all items. When the true data generating process is a GaPM-CDM, we set the weights to values such that $\sum_{k=1}^K \alpha_{jk} q_{jk} = 1$ for all items. For the IRFs, we assume the true $g_{jk}$s to be one of the Beta($3,3$), Beta($\frac{1}{3},\frac{1}{3}$), Beta($1,3$), or Beta($3,1$) cumulative distribution functions.

\subsubsection{Evaluation Criteria} 

For each simulated dataset, we fit the aPM-CDM and the GaPM-CDM and compare them in terms of parameter estimates, IRFs, factor score recovery, and model fit on unobserved data.

When the fitted model matches the true model (i.e., no model misspecification), we assess parameter recovery using the mean squared error (MSE). Let $\vartheta \in \bm{\Theta}_{\bm\kappa}$ denote a generic parameter under the GaPM-CDM and $\hat{\vartheta}^{(r)} \in \hat{\bm{\Theta}}_{\bm\kappa}^{(r)}$ its estimate for the $r$-th replication. For the aPM-CDM, we drop the subscript and write $\bm\Theta$ and $\hat{\bm\Theta}^{(r)}$ for the true and estimated parameters, respectively. The MSE is calculated as $\mathrm{MSE}(\hat{\vartheta}) = \frac{1}{R} \sum_{r=1}^R (\hat{\vartheta}^{(r)} - \vartheta)^2$. For simplicity, we report the average MSE (AvMSE) across all items for $\hat{\delta}_{jk}$s in the aPM-CDM and $\hat{\alpha}_{jk}$s in the GaPM-CDM such that $q_{jk} = 1$ in the \textbf{Q}-matrix, along with the AvMSE for the vector of free parameters in $\hat{\bm\Sigma} = \hat{\mathbf{L}} \hat{\mathbf{L}}^\top$. For the aPM-CDM, we also report the AvMSE for the estimated means $\hat{\bm\mu}$.

To evaluate IRF accuracy, we compute the integrated squared error (ISE):
\begin{equation*}
\mathrm{ISE}(\hat{\pi}_{j}^{(r)}) = \int_{[0,1]^K} \left( \hat\pi_j^{(r)}(\mathbf{U}) - \pi_j^{\mathcal{M}}(\mathbf{U}) \right)^2 \mathrm{d}\mathbf{U}\,, \qquad j = 1,\ldots,J,
\end{equation*}
where $\hat\pi_j^{(r)}(\mathbf{U})$ denotes the estimated IRF for item $j$ at replication $r$, and $\pi_{j}^\mathcal{M}(\mathbf{U})$ is the corresponding true IRF under model $\mathcal{M} \in$ \{aPM-CDM, GaPM-CDM\}. When the estimated model is the GaPM-CDM, we write ISE$_{\bm\kappa}(\hat\pi_j^{(r)})$ to emphasize the dependence of the approximation on the grid points. We report the average ISE (AvISE) across items and replications.

Lastly, for each $k = 1,\ldots,K$, we compare latent attribute recovery by computing Spearman's rank correlation between the EAP scores from the $r$-th replication and the true scores. We denote this correlation by $C_{k}$. The average rank correlation (AvC) across latent attributes and replications is reported.

To compare model fit, we compute the difference between marginal log-likelihoods of the GaPM-CDM and aPM-CDM evaluated on data not used in the estimation process. More specifically, after computing $\hat{\bm{\Theta}}_{\bm\kappa}^{(r)}$ for the GaPM-CDM and $\hat{\bm{\Theta}}^{(r)}$ for the aPM-CDM, we generate $500$ new observations from the true model and evaluate the corresponding marginal log-likelihoods $\ell(\hat{\bm{\Theta}}_{\bm\kappa}^{(r)})$ and $\ell(\hat{\bm{\Theta}}^{(r)})$ as in \eqref{eq:mllk(U)} but on the new sample. An importance sampling approach for computing the marginal log-likelihood is described in the Online Supplementary Materials. The difference $D^{(r)} = \ell(\hat{\bm{\Theta}}_{\bm\kappa}^{(r)}) - \ell(\hat{\bm{\Theta}}^{(r)})$ is then used for model comparison. If $D^{(r)} > 0$, the GaPM-CDM has a better fit on new data at replication $r$ than the aPM-CDM. The opposite is true if $D^{(r)} < 0$, and $D^{(r)} \approx 0$ implies similar fit. We report the average marginal log-likelihood difference $\bar{D}$ across all replications. As discussed in Remark \ref{rmk:comparison}, we use this measure of model fit because the GaPM-CDM and the aPM-CDM are not directly comparable due to the complex parametrization of the GaPM-CDM.

\subsubsection{Results}

The aPM-CDM and GaPM-CDM were estimated using the SA-MD algorithm in Section \ref{sec:Computation}. When the data is generated from the aPM-CDM, in the GaPM-CDM we use knots $\bm\kappa_0 = (0.05,0.1,0.2,...,0.9,0.95)^\top$, for which the added grid points near the boundary allow for better approximation of IRFs with guessing and slipping probabilities. When the data follows a GaPM-CDM, we use a set of equally spaced knots $\bm\kappa_1 = (0.05,...,0.95)^\top$. Further details on the selection of tuning parameters and initial values are given in the Online Supplementary Materials. 

Table \ref{tab:S1_ResultsA} shows that when the true data generating process follows an aPM-CDM, the GaPM-CDM can still perform relatively well, not far behind the aPM-CDM. The AvISE for the GaPM-CDM is not much higher than that of the aPM-CDM, and, while the differences between marginal log-likelihoods on new data are on average negative across all simulation settings, the value of $\bar D$ is not large. Moreover, comparing the AvC across models suggests that the GaPM-CDM recovers latent attributes and individual rankings just as well as the aPM-CDM. We conclude that, under model misspecification, the GaPM-CDM has slightly less generalization power and is less efficient than the aPM-CDM in terms of IRF and latent attributes recovery, but the gap between models is not large.

\begin{table}[!tb]
\caption{Simulation Study I (true model aPM-CDM). The AvMSE, AvISE, and AvC have been multiplied by 100 to allow for better numerical comparison.}
\label{tab:S1_ResultsA}
\begin{tabular}{@{}ccccccccccc@{}} \toprule
& & & \multicolumn{7}{c}{Estimated Model} \\ \cmidrule(lr){4-10} 
& & & \multicolumn{5}{c}{aPM-CDM} & \multicolumn{2}{c}{GaPM-CDM} \\ \cmidrule(lr){4-8} \cmidrule(lr){9-10}
\multirow{2}{*}{$\sigma$} & \multirow{2}{*}{$K$} & \multirow{2}{*}{$N$} & \multicolumn{3}{c}{AvMSE} & \multirow{2}{*}{AvISE} & \multirow{2}{*}{AvC} & \multirow{2}{*}{AvISE$_{\bm\kappa}$} & \multirow{2}{*}{AvC$_{\bm\kappa}$} & \multirow{2}{*}{$\bar{D}$} \\ \cmidrule(lr){4-6}
& & & $\hat{\bm\delta}_j$ & $\hat\sigma_{kk'}$ & $\hat{\bm\mu}$ & \\ \midrule
\multirow{4}{*}{$0.0$} & \multirow{2}{*}{$3$} & 1000  & 0.48 & 13.19 & 1.90 & 0.16 & 76.12 & 0.40 & 75.60 & $-$4.25 \\ 
& & 3000 & 0.20 & 3.68 & 0.59 & 0.07 & 76.35 & 0.21 & 76.16 & $-$1.72 \\ \cmidrule(lr){3-11} 
& \multirow{2}{*}{$5$} & 1000 & 0.82 & 18.06 & 6.78 & 0.32 & 65.26 & 0.47 & 65.07 & $-$0.98 \\ 
& & 3000 & 0.31 & 4.12 & 2.63 & 0.14 & 66.09 & 0.26 & 66.01 & $-$0.18 \\  \midrule
\multirow{4}{*}{$0.7$} & \multirow{2}{*}{$3$} & 1000 & 0.89 & 10.05 & 1.13 & 0.25 & 83.30 & 0.50 & 82.90 & $-$7.79 \\ 
& & 3000 & 0.34 & 2.76 & 0.39 & 0.10 & 83.67 & 0.23 & 83.50 & $-$3.75 \\ \cmidrule(lr){3-11} 
& \multirow{2}{*}{$5$} & 1000 & 1.66 & 14.20 & 3.02 & 0.49 & 79.55 & 0.61 & 79.25 & $-$9.53 \\ 
& & 3000 & 0.59 & 4.11 & 0.81 & 0.17 & 80.34 & 0.27 & 80.23 & $-$2.97 \\ \bottomrule
\end{tabular}
\end{table}

However, when data are generated from a GaPM-CDM, the aPM-CDM is highly misspecified, as suggested by the results in Table \ref{tab:S1_ResultsB}. Comparison of the AvISE and AvC between the GaPM-CDM and aPM-CDM show how the latter fails to recover the true IRFs and the latent variables scores. Moreover, the values of $\bar{D}$ are significantly larger than zero, suggesting a high impact of model misspecification in the aPM-CDM on generalizability to out-of-sample data, particularly when the latent attributes are correlated.

\begin{table}[!tb]
\caption{Simulation Study I (true model GaPM-CDM). The AvMSE, AvISE, and AvC have been multiplied by 100 to allow for better numerical comparison.}
\label{tab:S1_ResultsB}
\begin{tabular}{@{}cccccccccc@{}} \toprule
& & & \multicolumn{7}{c}{Estimated Model} \\ \cmidrule(lr){4-9} 
& & & \multicolumn{2}{c}{aPM-CDM} & \multicolumn{4}{c}{GaPM-CDM} \\ \cmidrule(lr){4-5} \cmidrule(lr){6-9}
\multirow{2}{*}{$\sigma$} & \multirow{2}{*}{$K$} & \multirow{2}{*}{$N$} & \multirow{2}{*}{AvISE} & \multirow{2}{*}{AvC} & \multicolumn{2}{c}{AvMSE$_{\bm\kappa}$} & \multirow{2}{*}{AvISE$_{\bm\kappa}$} & \multirow{2}{*}{AvC$_{\bm\kappa}$} & \multirow{2}{*}{$\bar{D}$} \\ \cmidrule(lr){6-7}
& & & & & $\hat{\bm\alpha}_j$ & $\hat\sigma_{kk'}$ \\ \midrule
\multirow{4}{*}{$0.0$} & \multirow{2}{*}{$3$} & 1000 & 1.62 & 83.58 & 0.72 & 0.23 & 0.32 & 84.61 & 18.99 \\
& & 3000 & 1.59 & 83.70 & 0.49 & 0.07 & 0.15 & 84.92 & 23.94 \\ \cmidrule(lr){3-10} 
& \multirow{2}{*}{$5$} & 1000 & 1.60 & 76.33 & 0.66 & 0.32 & 0.35 & 77.55 & 13.90 \\
& & 3000 & 1.54 & 76.69 & 0.44 & 0.10 & 0.18 & 77.82 & 16.56 \\  \midrule
\multirow{4}{*}{$0.7$} & \multirow{2}{*}{$3$} & 1000 & 1.59 & 88.11 & 1.71 & 0.14 & 0.48 & 89.09 & 53.53 \\
& & 3000 & 1.48 & 88.29 & 0.83 & 0.04 & 0.20 & 89.42 & 60.89 \\ \cmidrule(lr){3-10} 
& \multirow{2}{*}{$5$} & 1000 & 1.96 & 83.76 & 2.10 & 0.31 & 0.57 & 85.92 & 51.26 \\
& & 3000 & 1.81 & 84.06 & 0.97 & 0.10 & 0.24 & 86.39 & 55.57 \\ \bottomrule
\end{tabular}
\end{table}

\subsection{Study II: Exploratory Setting} \label{sec:Sims_S2}

This simulation study extends the previous one by fitting exploratory GaPM-CDMs (i.e., assuming the \textbf{Q}-matrix is unknown) when the data is generated from the GaPM-CDM. Data generation details are the same as before. The simulation exercise consists of $R = 100$ replications.

\subsubsection{Evaluation Criteria}

We report the same evaluation criteria as before. This time, however, we do not consider the information encoded in the \textbf{Q}-matrix when computing such metrics. That is, the AvMSE is computed over all estimated weights $\hat{\bm\alpha}_j$ and not only over those indicated to be positive by the \textbf{Q}-matrix. Similarly, the AvISE includes the contribution of estimated functions $\hat{g}_{jk}^s$ for which $q_{jk} = 0$ in the true IRF. This also applies to the recovered factor scores, and thus a similar caveat holds for the AvC. Therefore, the AvMSE, AvISE, and AvC for the exploratory GaPM-CDMs should reflect the lack of information, if any, coming from not incorporating the design of the \textbf{Q}-matrix into the measurement model. The results discussed below hold up to a permutation of the ordering of the latent attributes, adjusting the columns of the estimated weights and the rows and columns of the latent variables correlation matrix accordingly.

\subsubsection{Results}

We fit two exploratory GaPM-CDMs on each simulated data set, one using knots $\bm\kappa_1$ from Study I and the other using a coarser set of evenly spaced knots $\bm{\kappa}_2 = (0.1,\ldots,0.9)^\top$. Results in Table \ref{tab:S2_Results} show that exploratory GaPM-CDMs do not perform significantly worse than the confirmatory model, even without the information from the \textbf{Q}-matrix. The AvMSE for weights and factor correlations is comparable across models, sometimes smaller in the exploratory models, and, according to the AvISE and AvC, there is little loss of information in exploratory settings for IRF and latent attribute recovery. The average difference between the marginal log-likelihoods on the test data is negative in most cases, but the gap closes as the sample size increases. Our simulation results align with the discussion in Remark \ref{rmk:identif} on GaPM-CDM identifiability in the exploratory case. We can recover IRFs and, more importantly, factor scores, when the \textbf{Q}-matrix is unavailable or unreliable.

\begin{table}[ht]
\centering
\caption{Simulation Study II. Rows denoted by C$_{\bm{\kappa}_\star}$ and E$_{\bm{\kappa}_\star}$ correspond to the results for the confirmatory and exploratory GaPM-CDMs with knots $\bm\kappa_\star$, respectively. The AvMSE, AvISE, and AvC have been multiplied by 100 to allow for better numerical comparison.}
\label{tab:S2_Results}
\begin{tabular}{@{}cccccccccc@{}} \toprule
& & \multicolumn{4}{c}{$\sigma = 0$} & \multicolumn{4}{c}{$\sigma = 0.7$} \\ \cmidrule(lr){3-6} \cmidrule(lr){7-10}
& & \multicolumn{2}{c}{$K = 3$} & \multicolumn{2}{c}{$K = 5$} & \multicolumn{2}{c}{$K = 3$} & \multicolumn{2}{c}{$K = 5$} \\ \cmidrule(lr){3-4} \cmidrule(lr){5-6} \cmidrule(lr){7-8} \cmidrule(lr){9-10}
& $N$ & $1000$ & $3000$ & $1000$ & $3000$ & $1000$ & $3000$ & $1000$ & $3000$ \\ \midrule
\multirow{3}{*}{\shortstack{\\AvMSE$_{\bm\kappa}$\\$(\hat{\bm\alpha}_j)$}} & C$_{\bm{\kappa}_1}$ & 0.72 & 0.49 & 0.66 & 0.44 & 1.71 & 0.83 & 2.10 & 0.97 \\ 
& E$_{\bm{\kappa}_1}$ & 0.61 & 0.39 & 0.56 & 0.36 & 1.68 & 0.85 & 3.30 & 0.96 \\
& E$_{\bm{\kappa}_2}$ & 0.60 & 0.39 & 0.56 & 0.35 & 1.65 & 0.84 & 3.30 & 0.94 \\ \cmidrule(lr){2-10}
\multirow{3}{*}{\shortstack{\\AvMSE$_{\bm\kappa}$\\$(\hat{\sigma}_{kk'})$}} & C$_{\bm{\kappa}_1}$ & 0.23 & 0.07 & 0.32 & 0.10 & 0.14 & 0.04 & 0.31 & 0.10 \\
& E$_{\bm{\kappa}_1}$ & 0.35 & 0.16 & 0.57 & 0.27 & 0.82 & 0.49 & 2.50 & 2.07 \\
& E$_{\bm{\kappa}_2}$ & 0.36 & 0.17 & 0.57 & 0.27 & 0.95 & 0.59 & 2.86 & 2.20 \\ \cmidrule(lr){2-10}
\multirow{3}{*}{AvISE$_{\bm\kappa}$} & C$_{\bm{\kappa}_1}$ & 0.32 & 0.15 & 0.35 & 0.18 & 0.48 & 0.20 & 0.57 & 0.24 \\
& E$_{\bm{\kappa}_1}$ & 0.38 & 0.19 & 0.54 & 0.34 & 0.65 & 0.30 & 1.95 & 0.58 \\
& E$_{\bm{\kappa}_2}$ & 0.40 & 0.20 & 0.56 & 0.35 & 0.67 & 0.31 & 1.99 & 0.59 \\ \cmidrule(lr){2-10}
\multirow{3}{*}{AvC$_{\bm\kappa}$} & C$_{\bm{\kappa}_1}$ & 84.61 & 84.92 & 77.55 & 77.82 & 89.09 & 89.42 & 85.92 & 86.39 \\
& E$_{\bm{\kappa}_1}$ & 84.39 & 84.80 & 76.99 & 77.54 & 88.74 & 89.25 & 83.54 & 85.56 \\
& E$_{\bm{\kappa}_2}$ & 84.41 & 84.79 & 77.01 & 77.54 & 88.72 & 89.22 & 83.34 & 85.50 \\ \cmidrule(lr){2-10}
\multirow{2}{*}{\shortstack{\\$\bar{D}$\\$(\mathrm{C}_{\bm\kappa_1}~ \mathrm{vs.})$}} & E$_{\bm{\kappa}_1}$ & $-$2.74 & $-$1.10 & $-$6.71 & $-$2.27 & $-$1.51 & 0.08 & $-$14.69 & $-$2.80 \\
& E$_{\bm{\kappa}_2}$  & $-$2.51 & $-$1.00 & $-$6.47 & $-$2.20 & $-$1.22 & $-$0.07 & $-$14.92 & -2.66 \\ \bottomrule
\end{tabular}
\end{table}

\section{Real Data Applications} \label{sec:App}

We present results of the GaPM-CDM for the two motivating case studies. The first one is confirmatory, using the ECPE dataset introduced in Section \ref{sec:Data_ECPE}. The second one is exploratory, using PROMIS data from Section \ref{sec:Data_PROMIS}.

\subsection{English Test Data} \label{sec:App_ECPE}

The ECPE dataset contains responses for a sample of $N = 2922$ individuals to $J = 28$ items evaluating three ($K = 3$) latent attributes, namely knowledge of morphosyntactic rules (\textit{Morph.}), cohesive rules (\textit{Cohes.}), and lexical rules (\textit{Lexic.}) of the English language grammar. According to the \textbf{Q}-matrix, which can be found in \citet{ShangEtAl_AoAS2021}, 19 items measure a single attribute and 9 measure two attributes. No item requires all three attributes. In the current analysis, the \textbf{Q}-matrix is assumed to be known and incorporated in the models.

We apply the GaPM-CDM, aPM-CDM, ACDM, and GDINA models to the data. For the GaPM-CDM, we use knots $\bm{\kappa}_1 = (0.025, 0.05, 0.1, 0.2, \ldots, 0.9, 0.95, 0.975)^\top$, where the additional knots near the boundaries allow for modeling IRFs with guessing and slipping probabilities. Details on tuning parameters and initial values are discussed in the Online Supplementary Materials. As in Simulation Study I, we compare the generalization power of these models by computing the difference between marginal log-likelihoods on unobserved data. In this case, we perform a cross-validation (CV) exercise in which the observed data is randomly split into training (80\%) and testing sets (20\%). At the $r$-th CV replication, we fit the models to the training data and evaluate the marginal log-likelihood on the test data. We then compute the difference $D^{(r)}$, as defined earlier, for $r = 1,\ldots, 100$. Model comparison is based on the average difference $\bar{D}$ across all CV replications. The average marginal log-likelihood difference between the GaPM-CDM and the aPM-CDM is $\bar{D} = 1.99$ (with $(0.60, 3.37)$ as the 95\% confidence interval), showing that the GaPM-CDM with knots near the boundary performs better than the aPM-CDM in terms of out-of-sample prediction.

\begin{figure}[!ht]
\vspace{5mm}
    \centering
    \begin{subfigure}[t]{\textwidth}
        \centering
        \includegraphics[scale = 0.5]{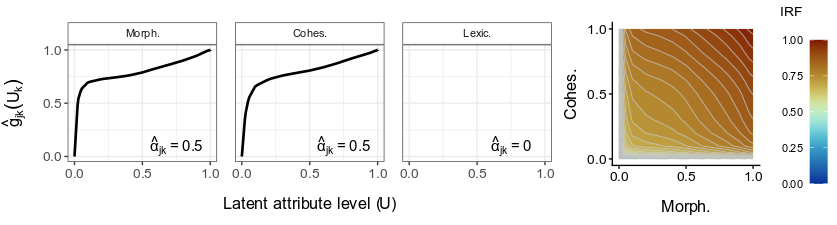}
        \caption{Item 1}
        \label{fig:ECPE_fitted_probs_a}
    \end{subfigure}
    \\
    \begin{subfigure}[t]{\textwidth}
        \centering
        \includegraphics[scale = 0.5]{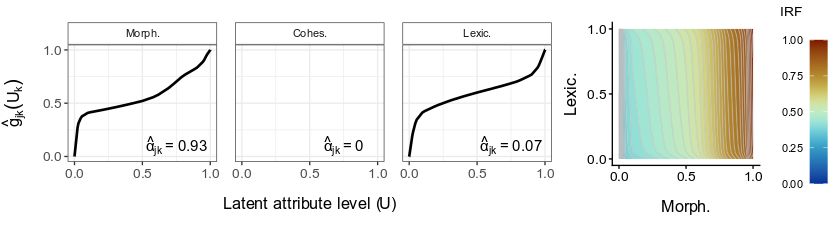}
        \caption{Item 3}
        \label{fig:ECPE_fitted_probs_b}
    \end{subfigure}
    \\
    \begin{subfigure}[t]{\textwidth}
        \centering
        \includegraphics[scale = 0.5]{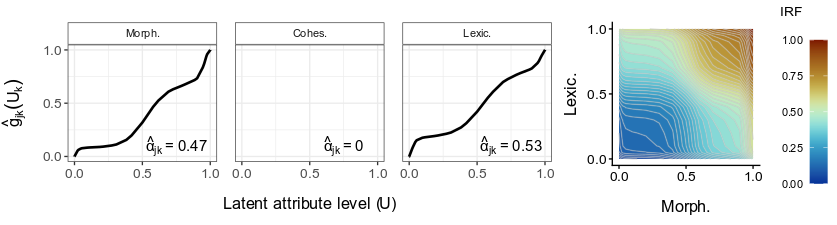}
        \caption{Item 12}
        \label{fig:ECPE_fitted_probs_c}
    \end{subfigure}
     \caption{Estimated weights ($\hat{\alpha}_{jk}$), estimated monotone functions ($\hat{g}_{jk}(U_k)$), and estimated IRF surface ($\hat{\pi}_j(U)$) under the GaPM-CDM, for selected items 1, 3, and 12 in the ECPE dataset.}
    \label{fig:ECPE_fitted_probs}
\end{figure}

In what follows, we discuss the estimation results of the GaPM-CDM with knots $\bm\kappa_1$ on the full dataset. Figure \ref{fig:ECPE_fitted_probs} shows some examples of the estimated monotone functions $\hat{g}_{jk}(U_k)$ and the corresponding IRF surfaces $\hat{\pi}_j(U)$. These items were selected from those that require two latent attributes according to the \textbf{Q}-matrix. For instance, item 1 (Figure \ref{fig:ECPE_fitted_probs_a}) has a high guessing probability at the lower end of the \textit{Morph.} and \textit{Cohes.} scales, with comparable contributions from both attributes ($\hat{\alpha}_{1,\mathrm{M}}=0.50$ and $\hat{\alpha}_{1,\mathrm{C}}=0.50$). The IRF then increases in \textit{Morph.} and \textit{Cohes.}, with high probabilities of correct response across the full range of both attribute values.

The estimated functions $\hat{g}_{3,\mathrm{M}}$ and $\hat{g}_{3,\mathrm{L}}$ (Figure \ref{fig:ECPE_fitted_probs_b}) indicate that item 3 has high guessing and moderate slipping probabilities, consistent with benchmark model estimates. Unlike these models, the GaPM-CDM captures the non-linear relationships between latent attributes and response probabilities. The estimated weights ($\hat{\alpha}_{3,\mathrm{M}}=0.93$ and $\hat{\alpha}_{3,\mathrm{L}}=0.07$) show that the IRF surface $\hat{\pi}_3(U)$ is driven mainly by \textit{Morph.}, with low contribution from \textit{Lexic.}, further suggesting possible misspecification in the corresponding row of the \textbf{Q}-matrix. Item 12 (Figure \ref{fig:ECPE_fitted_probs_c}) shows low guessing and moderate slipping probabilities. The estimated functions $\hat{g}_{12,\mathrm{M}}$ and $\hat{g}_{12,\mathrm{L}}$ (and corresponding weights $\hat{\alpha}_{12,\mathrm{M}}=0.47$ and $\hat{\alpha}_{12,\mathrm{L}}=0.53$) yield a highly non-linear IRF surface with low response probabilities for \textit{Morph.} and \textit{Lexic.} below 0.5, rising above that threshold.

Table \ref{tab:A1_ECPE_Sigma} shows the estimated covariance and correlation matrices of the latent attributes for the aPM-CDM and the GaPM-CDM, respectively. The diagonal entries of the covariance matrix are close to one and the estimated means for the Gaussian copula in the aPM-CDM are all relatively small. For comparison, we also present the correlations for the aPM-CDM (above diagonal) and the means in the 0-1 scale. Figures showing the estimated EAP factor scores are included in the Online Supplementary Materials. The scatterplots for the aPM-CDM factor scores are consistent with those in \citet{ShangEtAl_AoAS2021}. While factor scores for the GaPM-CDM are constrained by the fixed means in the Gaussian copula, the factor scores produced by the two models are largely consistent in terms of their rank order, with Spearman's rank correlations of $0.99$ for all for three attributes.

\begin{table}[!htp]
\caption{Estimated covariance (for aPM-CDM) and correlation (for GaPM-CDM) matrices for the latent attributes. ECPE dataset. $^\dagger$Entries in italics above the diagonal are correlations in aPM-CDM. $^\ddagger$Underlined entries are fixed parameters in the GaPM-CDM.}
\label{tab:A1_ECPE_Sigma}
\begin{tabular}{@{}cccccccc@{}} \toprule
& \multicolumn{3}{c}{aPM-CDM$^\dagger$} & & \multicolumn{3}{c}{GaPM-CDM$^\ddagger$} \\ \cmidrule(lr){2-4} \cmidrule(lr){5-8} 
& Morph. & Cohes. & Lexic. & & Morph. & Cohes. & Lexic. \\ \midrule 
Morph. & 1.26 & \textit{0.79} & \textit{0.88} & & \underline{\textit{1.0}} & 0.83 & 0.86 \\
Cohes. & 0.92 & 1.09 & \textit{0.82} & & & \underline{\textit{1.0}} & 0.82\\
Lexic. & 0.98 & 0.85 & 0.97 & & & & \underline{\textit{1.0}} \\ \midrule
$\hat{\bm\mu}$ & $-$0.37 & 0.42 & 0.82 & & \underline{\textit{0}} & \underline{\textit{0}} & \underline{\textit{0}} \\
$\hat{\bm\mu}$ (0-1 scale) & 0.36 & 0.66 & 0.79 & & \underline{\textit{0.5}} & \underline{\textit{0.5}} & \underline{\textit{0.5}} \\ \bottomrule
\end{tabular}
\end{table}

\subsection{Patient-reported Outcomes} \label{sec:App_PROMIS}

We analyze a sub-sample from wave 1 of PROMIS, consisting of $N = 737$ non-clinical patients who responded to all the items. The median age for the respondents in our sample is $51$ ($\min = 18$, $\mathrm{q}_{25} = 39$, $\mathrm{q}_{75} = 64$, $\max = 87$), and sex was evenly distributed ($49.9\%$ male, $50.1\%$ female). Observed responses were originally measured on a 5-point Likert scale and were positively oriented (i.e., higher values mean better social role performance, SRPPER). However, for the purpose of this paper, we dichotomized the original responses.\footnote{Values greater than or equal to 4 in the original 5-point Likert scale were assigned a value of 1, and 0 otherwise. This cut-off point produced items with balanced classes. A robustness check using 3 as the cut-off point produced similar results when selecting $K$ but produced more imbalanced items.} We apply the GaPM-CDM and aPM-CDM to this dataset under an exploratory setting.

\begin{figure}[!ht]
\centering
\vspace{5mm}
\includegraphics[scale = 0.6]{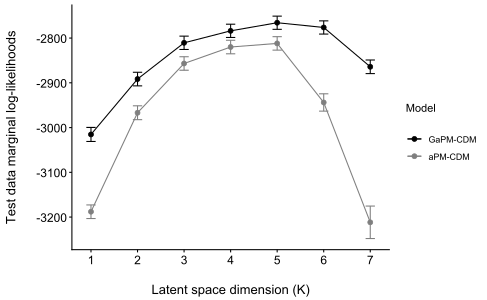}
\caption{Cross-validation test-data marginal log-likelihood for the GaPM-CDM and aPM-CDM. Error bars correspond to the standard error of the average test-data marginal log-likelihood for each level of $K = 1,\ldots,7$.}
\label{fig:A2_PROMIS_Kselect}
\end{figure}

We first learn the number of attributes using cross-validation, as follows. Let the number of latent attributes $K$ iterate over a set of candidate values $\mathbb K = \{1,\ldots,7\}$. For a given $K$, we randomly split the observed data into training (80\%) and testing (20\%) sets $r = 1,\ldots,R =100$ times. At each replication, we fit both aPM-CDM and GaPM-CDM on the training data to obtain the (S)MMLEs $\hat{\bm\Theta}^{(r)}$ and $\hat{\bm\Theta}_{\bm\kappa}^{(r)}$, respectively, and then we evaluate the marginal log-likelihood on the testing data. The cross-validation `error' for a given $K$ is then computed as $\mathrm{CVE}(K) = \frac{1}{R} \sum_{r = 1}^R \ell(\hat{\bm\Theta}^{(r)})$ for the aPM-CDM and $\mathrm{CVE}_{\bm\kappa}(K) = \frac{1}{R} \sum_{r = 1}^R \ell(\hat{\bm\Theta}^{(r)}_{\bm\kappa})$ for the GaPM-CDM. For each model, we select the latent dimension by $\hat K = \argmax_{K \in \mathbb K}\left\{ \mathrm{CVE}(K)\right\}$, giving $\hat{K} = 5$ in both cases (Figure \ref{fig:A2_PROMIS_Kselect}). An important result is that the GaPM-CDM fits the testing data better than the aPM-CDM across all potential dimensions of the latent attribute space. The small standard errors relative to the test-data log-likelihood values suggest model stability across folds and latent dimensionality $K$. For the selected $\hat K$, we fit the models to the whole dataset with knots $\bm{\kappa} = (0.05,\ldots,0.95)^\top$ and discuss results below. Implementation details and sensitivity analysis to the number of knots are reserved for the Online Supplementary Materials.

\begin{figure}[!ht]
\centering
\vspace{5mm}
\includegraphics[scale = 0.6]{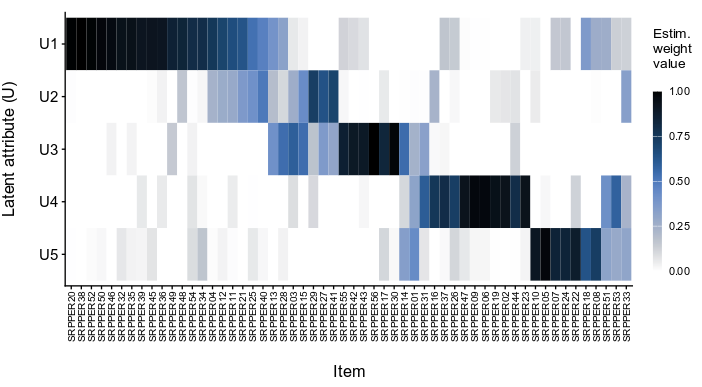}
\caption{Matrix of estimated weights (transposed) for the $\hat{K} = 5$ dimensional GaPM-CDM on the PROMIS dataset.}
\label{fig:A2_PROMIS_Ahat}
\end{figure}

In the absence of a design \textbf{Q}-matrix, model interpretation is based on the recovered structure of the matrix of estimated weights with rows $\hat{\bm\alpha}_j^\top$ for all 56 items and the estimated correlation matrix in the Gaussian copula model for the five latent attributes. Figure \ref{fig:A2_PROMIS_Ahat} shows a sparse (transpose) matrix of estimated weights which allows for clear interpretation of the recovered latent attributes. We implement an arbitrary permutation of the items and the latent attributes for better visualization. Most items mostly measure one of the latent attributes, while a small subset of them are related more than one.

The first attribute ($U_1$) is measured by items associated with \textit{personal and social leisure functioning}. Questions loading on this attribute relate to the patient's perceived capacity to engage in recreational and community activities, particularly with friends, and to fulfil social roles and responsibilities. Some examples of items with large contributions are SRPPER20 (\textit{``I am able to do all of the activities with friends that are really important to me''}), SRPPER45 (\textit{``I can keep up with my social responsibilities''}), and SRPPER50 (\textit{``I am able to do all of the community activities that are really important to me''}).

The second latent attribute ($U_2$) focuses on the individual's perceived \textit{personal and social leisure functioning capacity}. Items with large estimated weights on this attribute are mostly negatively worded questions on restrictions, constraints, or barriers that affect their degree of autonomy, freedom, and satisfaction about their ability to engage in personal leisure and recreational activities. Recall that negatively worded items have already been reverse-coded in the data, meaning that high scores on these items are associated with higher personal and social leisure functioning. However, the emergence of two attributes ($U_1$ and $U_2$) about essentially the same construct of social function reflects the nuance of item wording. Some examples of items measuring this latent attribute are SRPPER29 (\textit{``I have to do my hobbies or leisure activities for shorter periods of time than usual (Reversed)''}) and SRPPER41 (\textit{``I have to limit my hobbies or leisure activities (Reversed)''}).

The third recovered latent attribute ($U_3$) is related to the individual's perceived \textit{family and household role functioning capacity}. Items loading on this latent attribute are negatively worded questions on the person's restrictions in fulfilling personal and domestic roles, including time and capacity limitations. As before, items were reverse-coded so higher scores on these items reflect higher family and household role functioning. Note how some items load similarly on $U_2$ and $U_3$, revealing close similarities between latent constructs and reflecting the nuance of item wording. Some examples include items SRPPER30 (\textit{``I feel limited in my ability to visit relatives (Reversed)''}), SRPPER56 (\textit{``I feel limited in the amount of time I have to visit relatives (Reversed)''}).

The fourth latent attribute ($U_4$) can be interpreted as the individual's \textit{work functioning}, broadly defined to include both formal employment and domestic labor. Items weighing heavily on this dimension reflect the individual's perceived ability and limitations in meeting work-related expectations and responsibilities. Some examples include items SRPPER06 (\textit{``I am accomplishing as much as usual at work, including work at home''}), SRPPER23 (\textit{``I am able to do all of my usual work, including work at home''}), and SRPPER47 (\textit{``I can keep up with my work responsibilities, including work at home''}).

The underlying concept of the fifth latent attribute ($U_5$) is \textit{family role functioning}. This dimension concerns the individual's perceived ability to engage in and fulfil family-related obligations and activities. Items with large estimated weights include SRPPER10 (\textit{``I am able to do all of my regular family activities''}) and SRPPER05 (\textit{``I can do everything for my family that I feel I should do''}). Similarly, the presence of two attributes ($U_3$ and $U_5$) that essentially describe the same construct, family role function, reflects the influence of item wording.

The recovered latent attributes are highly correlated, as reported in Table \ref{tab:A2_PROMIS_Sigmahat}. This explains why some items load on more than one attribute. Indeed, closer inspection reveals how these items involve social role performance in areas that overlap with the identified attributes. For example, item SRPPER33 (\textit{``I am able to run errands as much as usual''}) loads on \textit{personal and social leisure} ($U_1$ and $U_2$), \textit{work} ($U_4$), and \textit{family} ($U_5$) functioning capacity; and item SRPPER53 (\textit{``I am able to do all of the work that people expect me to do, including work at home''}) measures the individuals'  \textit{work} ($U_4$) and \textit{family role} ($U_5$) functioning, while also tapping into the \textit{personal and social leisure functioning} ($U_1$) dimension.

\begin{table}[!htp]
\caption{Estimated correlation matrix for the five recovered latent attributes. PROMIS dataset. Underlined entries are fixed parameters in the GaPM-CDM.}
\label{tab:A2_PROMIS_Sigmahat}
\begin{tabular}{@{}cccccc@{}} \toprule
& $U_1$ & $U_2$ & $U_3$ & $U_4$ & $U_5$ \\ \midrule 
$U_1$ & \underline{\textit{1.0}} & 0.78 & 0.76 & 0.75 & 0.86 \\ 
$U_2$ & & \underline{\textit{1.0}} & 0.76 & 0.73 & 0.69 \\ 
$U_3$ & & & \underline{\textit{1.0}} & 0.61 & 0.68 \\ 
$U_4$ & & & & \underline{\textit{1.0}} & 0.83 \\ 
$U_5$ & & & & & \underline{\textit{1.0}} \\ \bottomrule
\end{tabular}
\end{table}

\section{Discussions} \label{sec:Disc}

This paper presents a semiparametric extension of the PM-CDM proposed by \cite{ShangEtAl_AoAS2021} and further develops a sieve-based estimator for model estimation and a stochastic optimization method for its computation. Simulation studies and two real data examples demonstrate that the proposed model is more flexible than its parametric counterpart and yields better fits to the data. In particular, Simulation Study II and the application to PROMIS data suggest that the GaPM-CDM is identifiable in exploratory settings with no prior knowledge of the \textbf{Q}-matrix of the item-attribute relationship, thereby substantially expanding the scope of PM-CDMs in their applications.

We focus on binary item responses. However, other response types may appear in cognitive diagnosis, including ordinal responses, continuous responses such as response times, and count-valued responses from tests with repetitive tasks or eye-tracking sensors. We believe that a general framework of GaPM-CDMs can be developed for different types of multivariate responses by extending the parametric framework of \cite{Lee&Gu_Psychometrika2024}. For example, for ordinal responses, we can model the conditional probabilities of adjacent categories with the monotone generalized additive form that we introduced in the current model. 

Many CDMs and PM-CDMs model interactions among attributes, which may be useful for uncovering the complex psychological processes underlying item responses. We anticipate that it is possible to incorporate interaction terms semiparametrically into the current framework. However, the identifiability of such a model is more complicated, and additional constraints are required on the model parameters during parameter estimation. We leave it for future investigation. 

Finally, some theoretical properties of the GaPM-CDM still need to be established. In particular, we provided only sufficient conditions for generic identifiability and discussed heuristic guarantees of the estimator's consistency. As discussed earlier, conditions for model identifiability and additional consistency results can be derived only under the nontrivial double-asymptotic regime, where both $J$ and $N$ grow to infinity. Several challenges arise in this setting. First, the dimension of the parameter space grows together with $J$, for which many asymptotic tools are not directly applicable. Second, the marginal likelihood is difficult to analyze, as it involves integrals that are analytically intractable. Third, the model involves infinite-dimensional functions that are approximated by a set of basis functions. Fourth, the true parameters may lie on the boundaries of the parameter space. Deriving a new information criterion with a penalty function that accounts for the dimensionality of the sieve approximations and their bias is also of interest, but left for future investigation.

\bibliographystyle{apalike}
\bibliography{bib_GaPM-CDM}

\end{document}


\maketitle

\setcounter{section}{0}
\setcounter{equation}{0}
\setcounter{table}{0}

\renewcommand{\thesection}{\Alph{section}}
\renewcommand{\thesubsection}{\thesection\arabic{subsection}}
\renewcommand{\theequation}{\thesection.\arabic{equation}}
\renewcommand{\thetable}{A\arabic{table}}
\renewcommand{\thefigure}{A\arabic{figure}}

\counterwithin*{equation}{section}

This supplementary material includes a discussion on the boundary behavior of the GaPM-CDM IRFs in Section \ref{sec:OSM_bound}, details for the proposed stochastic-approximation mirror descent (SA-MD) algorithm in Section \ref{sec:OSM_SA-MD}, details on the importance sampling algorithm used for computing the marginal log-likelihood in Section \ref{sec:OSM_mllk+IS}, implementation details of the simulation studies in Section \ref{sec:OSM_Simulation}, and further results for the empirical applications in Section \ref{sec:OSM_Apps}. Relevant theoretical results are included in Section \ref{sec:OSM_proof}.

\section{On the boundary behavior of the GaPM-CDM IRFs} \label{sec:OSM_bound}

Recall from Section 3.2 that we impose boundary conditions $g_{jk}(0) = 0$ and $g_{jk}(1) = 1$ on the nonparametric functions in the additive representation of the IRF $\pi_j(\mathbf{U})$ for identifiability. Moreover, as one of the reviewers correctly pointed out, realizations of the latent attributes in the boundaries (i.e., $\{U_k = 0\}$ and $\{U_k = 1\}$ for $k = 1,\ldots,K$) are measure-zero events in the (Ga)PM-CDM framework. This representation effectively precludes strictly positive guessing (i.e., $\pi_j(\mathbf{U})>0$ for $U_k = 0$ when $q_{jk} = 1$) and slipping (i.e., $1-\pi_j(\mathbf{U}) > 0$ for $U_k = 1$ when $q_{jk} = 1$). However, the nonparametric functions $g_{jk}$ can approximate near-guessing and near-slipping behaviors when evaluated at a point $\mathbf{U} \in (0,1)^K$ that is arbitrarily close to the boundary.

\begin{figure}[!ht]
        \centering
        \vspace{5mm}
        \includegraphics[scale = 0.65]{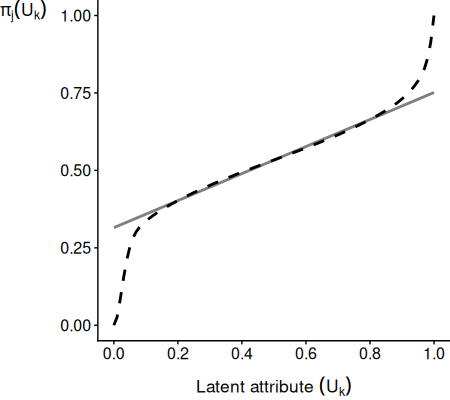}
        \caption{Example of how an aPM-CDM IRF for an item with guessing $(\delta_{j0} > 0)$ and slipping $(1-\delta_{j0}-\delta_{jk}> 0)$ parameters (solid line, \solid) is approximated by an appropriately chosen smooth function $g_{jk}(U_k)$ (dashed line, \dashed) in the GaPM-CDM.}
        \label{fig:IRF_approximation}
    \end{figure}

More precisely, for any $\epsilon > 0$, an appropriate choice of the $g_{jk}$s in the IRF can represent any guessing probability $G > 0$ such that $\pi_j(\mathbf{U}) \leq G$ for all relevant $\mathbf{U} \in [0, \epsilon]^K$, and any slipping probability $S> 0$ such that $1 - \pi_j(\mathbf{U}) \leq S$ for all relevant $\mathbf{U} \in [1-\epsilon, 1]^K$. An example is depicted in Figure \ref{fig:IRF_approximation}, where a well-calibrated $g_{jk}$ approximates $\pi_j(U_k) > 0$ for some $U_k \in [0,\epsilon]$ and $1-\pi_j(U_k) > 0$ for the some $U_k \in [1-\epsilon,1]$.

\section{Computational and implementation details of the SA-MD algorithm} \label{sec:OSM_SA-MD}

In what follows, we elaborate further on the stochastic approximation (SA) and mirror descent (MD) steps. Let $t$ denote the iteration number and $\bm\Theta_{\bm \kappa}^{(t)}$ be the parameter value at the $t$-th iteration. We then discuss in detail the Metropolis-Adjusted Langevin algorithm (MALA) and the analytical forms of the MD update rules.

\subsection{SA step} \label{sec:OSM_SA}

In the $t$-th iteration, the SA step constructs an approximate stochastic gradient for $\ell(\bm\Theta_{\bm \kappa})$, which takes the form of
\begin{equation}
G(\bm\Theta_{\bm \kappa}^{(t-1)}, \mathbf{U}_1^{(t)},\ldots, \mathbf{U}_N^{(t)}) = \sum\limits_{i = 1}^{N} \nabla_{\bm{\Theta}_{\bm{\kappa}}}\, \log f(\mathbf{y}_i, \mathbf{U}^{(t)}_{i}; \bm{\Theta}_{\bm\kappa})\lvert_{\bm{\Theta}_{\bm\kappa} = \bm{\Theta}_{\bm\kappa}^{(t-1)}} \,,  \label{eq:stochG}
\end{equation}
where, for an individual $i$, $\mathbf y_i = (y_{ij}: j = 1,\ldots,J)^\top$ is the vector of observed responses, $f(\mathbf{y}_i,\mathbf{U}_i)$ is the complete-data density function, and $\mathbf U_i^{(t)}$ is an approximate sample of the partial-mastery scores under the posterior density
\begin{align}\label{eq:poster}
f(\mathbf U \cond \mathbf{y}_i; \bm{\Theta}_{\bm\kappa}^{(t-1)}) \propto & ~ f(\mathbf y_i, \mathbf U; \bm \Theta_{\bm\kappa}^{(t-1)} ) \nonumber \\[0.25em]
= & ~\left[\prod\limits_{j = 1}^J \pi_{j}(\mathbf{U}; \bm{\alpha}_j^{(t-1)}, \bm{\theta}_{j}^{(t-1)}, \bm{\kappa})^{y_{ij}} \, (1-\pi_{j}(\mathbf{U}; \bm{\alpha}_j^{(t-1)}, \bm{\theta}_{j}^{(t-1)}, \bm{\kappa}))^{1-y_{ij}}\right] \times \nonumber \\[0.25em]
& ~ \mathrm{d}D(\mathbf{U}; \mathbf{L}^{(t-1)})\,.
\end{align}
It can be shown that, when $\mathbf U_i^{(t)}$ is an exact sample from \eqref{eq:poster}, then $G(\bm\Theta_{\bm \kappa}^{(t-1)}, \mathbf{U}_1^{(t)}, \ldots, \mathbf{U}_N^{(t)})$ is an exact stochastic gradient of $\ell(\bm\Theta_{\bm \kappa})$, in the sense that 
\begin{equation*}
\mathbb{E} \left( G(\bm\Theta_{\bm \kappa}^{(t-1)}, \mathbf U_1^{(t)}, \ldots, \mathbf U_N^{(t)}) \right) = \left. \frac{\partial\ell(\bm\Theta_{\bm \kappa})}{\partial \bm\Theta_{\bm \kappa}} \right|_{\bm\Theta_{\bm \kappa} = \bm\Theta_{\bm \kappa}^{(t-1)}}\,,
\end{equation*}
where the expectation is with respect to $\mathbf U_1^{(t)}, \ldots, \mathbf U_N^{(t)}$ under the posterior distribution \eqref{eq:poster}.

Since the posterior distribution \eqref{eq:poster} has a relatively complex form, sampling exactly from this distribution is difficult. In our implementation, we perform MCMC sampling of the latent attributes from the posterior distribution based on the Metropolis-Adjusted Langevin algorithm (MALA, \citealp{Roberts&Tweedie_Bernoulli1996, Oliviero-Durmus&Moulines_Biometrika2024}), following its computational efficiency and fast convergence in high dimensional settings \citep{Roberts&Rosenthal_JRSSB1998}.

\subsection{MD step} \label{sec:OSM_MD}

The MD step is a stochastic version of the mirror descent algorithm \citep{Beck&Teboulle_ORL2003} in which the model parameters are updated based on the stochastic gradient in \eqref{eq:stochG}. The MD update rule takes the form
\begin{equation}
\bm{\Theta}_{\bm\kappa}^{(t)} = \argmin_{\bm{\Theta}_{\bm\kappa} \in \bm{\Xi}_{\bm\kappa} } \left\{ G(\bm\Theta_{\bm \kappa}^{(t-1)}, \mathbf{U}_1^{(t)}, \ldots, \mathbf{U}_N^{(t)})^\top \bm{\Theta}_{\bm\kappa} + \dfrac{1}{\gamma^{(t)}} D_{\psi}(\bm\Theta_{\bm\kappa},\bm\Theta_{\bm\kappa}^{(t-1)})\right\}\,, \label{eq:Theta[t+1]_SMD}
\end{equation}
where $\gamma^{(t)} > 0$ is a decaying step size and
\begin{equation*}
D_\psi(\bm\Theta_{\bm\kappa},\bm\Theta_{\bm\kappa}^{(t-1)}) = \psi(\bm\Theta_{\bm\kappa}) - \psi(\bm\Theta_{\bm\kappa}^{(t-1)}) - (\bm\Theta_{\bm\kappa}-\bm\Theta_{\bm\kappa}^{(t-1)})^\top\nabla_{\bm\Theta_{\bm\kappa}}\psi(\bm\Theta_{\bm\kappa}^{(t-1)})
\end{equation*}
is known as the Bregman divergence based on the distance-generating function
\begin{equation*}
\psi(\bm \Theta_{\bm\kappa}) = \sum\limits_{j = 1}^J \sum\limits_{k = 1}^K \alpha_{jk} \log(\alpha_{jk}) + \sum\limits_{j = 1}^J \sum\limits_{k = 1}^K \sum\limits_{l = 1}^{L+1} \theta_{jk,l} \log(\theta_{jk,l}) + \frac{1}{2}\sum\limits_{k = 1}^K ||\bm l_k||^2_2  \label{eq:psi_dgf}
\end{equation*}

The Bregman divergence in \eqref{eq:Theta[t+1]_SMD} acts as a penalty term that prevents the update rule from producing parameters that are `too far' from the current point based on some distance determined by $\psi$, which is adapted to the local geometry of the parameter space. In particular, the weights $\bm\alpha_j$ and sieve approximation parameters $\bm\theta_{jk}$ live in the $K$- and $ (L+1)$-dimensional probability simplexes, respectively, and therefore we use the negative entropy to derive this distance. In this case, the update for these parameters in \eqref{eq:Theta[t+1]_SMD} has a closed analytic form, which can be easily computed. For the parameters in the $k$-th row of $\mathbf{L}$ we use the Euclidean distance and thus \eqref{eq:Theta[t+1]_SMD} becomes the proximal operator used in the proximal stochastic gradient descent \citep{Zhang&Chen_Psychometrika2022, AtchadeEtAl_JMLR2017}. In fact, since each row $\bm l^\top_k$ in $\mathbf L$ satisfies $||\bm l_k||_2^2 = 1$, the update for $\bm l_k$ becomes a projected gradient descent update, which has an analytic closed form solution. Consequently, the MD step \eqref{eq:Theta[t+1]_SMD} can be computed with a low computational cost.

Convergence of the sequence $\bm{\Theta}_{\bm{\kappa}}^{(t)}$, $t = 1,2,\ldots$, to a stationary point $\bm\Theta^*_{\bm\kappa} \in \bm\Xi_{\bm\kappa}$ depends on an appropriate choice of the step sizes such that $\lim_{t \to \infty} \sum_t (\gamma^{(t)})^2 < \infty$ and $\lim_{t \to \infty} \sum_t \gamma^{(t)} = \infty$ \citep{Robbins&Monro_AoMS1951}. Following theoretical results in \citet{Zhang&Chen_Psychometrika2022}, our choice of step size is $\gamma^{(t)} = \mu\, t^{-0.5-\epsilon}$ for some constant $\mu > 0$ and arbitrarily small $\epsilon > 0$. To improve the performance of the proposed algorithm, we can let $\mu$ depend on the specific type of parameter in $\bm\Theta_{\bm\kappa}$.

We further comment on the implementation of the SA-MD algorithm. First, our stopping criteria are based on a fixed but large number of iterations $T$. This avoids bias from premature stopping and better aligns with the convergence of stochastic approximation methods as $T \to \infty$. Second, we compute the SMMLE as the Polyak-Ruppert trajectory average \citep{Polyak&Juditsky_SIAM1992, Ruppert_TechRep1988} of the parameter updates at iterations beyond a fixed burn-in period $\omega < T$. That is, $\hat{\bm\Theta}_{\bm\kappa} = (\sum_{t = \omega+1}^T\bm\Theta^{(t)}_{\bm\kappa})/(T-\omega)$. This approach improves convergence speed and stability of the SMMLE and reduces sensitivity to noisy updates. The pseudo-code of the estimation procedure is described in Algorithm \ref{alg:SA-MD} and it has been implemented in the \texttt{R} package \texttt{gapmCDM}, available online at \href{https://github.com/ccardehu/pmCDM}{https://github.com/ccardehu/gapmCDM}.

\begin{algorithm}
\caption{Pseudo-code for SA-MD} \label{alg:SA-MD}
\begin{algorithmic}
\medskip
\Statex \textbf{Settings:} Iteration limit $T$, burn-in period $\omega$, step-size constant $\mu > 0$.
\Statex \textbf{Input:} Complete data $(\mathbf{y}_i, \mathbf{U}_i^{(0)}: i = 1,\ldots,N)$ and starting values $\bm\Theta^{(0)}$.
\For{$t = 1,\ldots,T$}
\Statex SA step, eqs. \eqref{eq:stochG} and \eqref{eq:poster}:
\For{$i = 1,\ldots,N$}
\State Sample: $\mathbf{U}_i^{(t)} \sim f(\mathbf U \cond \mathbf{y}_i; \bm{\Theta}_{\bm\kappa}^{(t-1)} )$
\EndFor
\State Compute: $G(\bm\Theta_{\bm \kappa}^{(t-1)}, \mathbf{U}_1^{(t)}, \ldots, \mathbf{U}_N^{(t)}) = \sum\limits_{i = 1}^{N} \nabla_{\bm{\Theta}_{\bm{\kappa}}}\, \log f(\mathbf{y}_i, \mathbf{U}^{(t)}_{i}; \bm{\Theta}_{\bm\kappa}^{(t-1)})$
\Statex MD step, eq. \eqref{eq:Theta[t+1]_SMD}:
\State Update: $\bm{\Theta}_{\bm\kappa}^{(t)} = \argmin_{\bm{\Theta}_{\bm\kappa} \in \bm{\Xi}_{\bm\kappa} } \left\{ G(\bm\Theta_{\bm \kappa}^{(t-1)}, \mathbf{U}_1^{(t)}, \ldots, \mathbf{U}_N^{(t)})^\top \bm{\Theta}_{\bm\kappa} + \dfrac{1}{\gamma^{(t)}} D_{\psi}(\bm\Theta_{\bm\kappa},\bm\Theta_{\bm\kappa}^{(t-1)})\right\}$
\EndFor
\Statex \textbf{Output:} $\hat{\bm\Theta}_{\bm\kappa} = (\sum_{t = \omega+1}^T\bm\Theta^{(t)}_{\bm\kappa})/(T-\omega)$
\end{algorithmic}
\end{algorithm}

\subsection{Metropolis-Adjusted Langevin Algorithm (MALA)} \label{sec:OSM_MALA}

For computational simplicity, we formulate the GaPM-CDM in terms of the transformed latent attributes $\mathbf{Z} = (Z_k: k = 1, \ldots, K)^\top$, with $Z_k = \Phi^{-1}(U_k)$. The multidimensional integral in the marginal log-likelihood thus is taken over $\R^K$ with respect to the measure defined by the $K$-variate Normal distribution $f(\mathbf{Z};\bm\Theta_{\bm\kappa}) \sim \mathrm{N}_K(\bm{0}, \bm{\Sigma})$. The complete-data and conditional density functions in \eqref{eq:stochG} and \eqref{eq:poster} are also expressed in terms of $\mathbf{Z}$.

A basic description of the Metropolis-Adjusted Langevin algorithm \citep[MALA,][]{Roberts&Tweedie_Bernoulli1996, Oliviero-Durmus&Moulines_Biometrika2024} is as follows. The MALA is an efficient MCMC sampler that requires minimal tuning in practice and has satisfactory empirical performance in high-dimensional settings. The MALA uses information from the local geometry of the target distribution, $f(\mathbf{Z}\cond \mathbf{y}_i;\bm\Theta_{\bm\kappa})$, to explore areas around the mode of such distribution faster and more efficiently than other MCMC samplers. The MALA is based on the Langevin dynamics solution of the following $K$-dimensional stochastic differential equation:
\begin{equation}
    \mathrm{d}\mathbf{Z}(t) = -\nabla_{\mathbf{Z}} \log f(\mathbf{y}_i, \mathbf{Z}; \bm\Theta_{\bm\kappa})\, \mathrm{d}t + \sqrt{2}\mathrm{d}\mathbf{B}(t)\,, \quad i = 1,\ldots,N\,, \label{eq:dZt}
\end{equation}
where $\mathbf{B}(t)$, $t = 1,2,\ldots$, is a standard $K$-dimensional Brownian motion. Under mild assumptions on $f(\mathbf{Z} \cond \mathbf{y}_i; \bm\Theta_{\bm\kappa})$, it is known that the continuous-time stochastic process $\mathbf{Z}(t)$ has a strong solution with $f(\mathbf{Z} \cond \mathbf{y}_i; \bm\Theta_{\bm\kappa})$ as its invariant probability measure \citep{Roberts&Tweedie_Bernoulli1996, Durmus&Moulines_AoAP2017}.

In practice, sampling $\mathbf{Z}_i^{(t+1)} \sim f(\mathbf{Z} \cond \mathbf{y}_i;\Theta^{(t)})$ from the continuous-time process is not feasible, and thus a discrete-time approximation is required. The Euler-Maruyama discretization of \eqref{eq:dZt} gives the following update rule:
\begin{align}
    \mathbf{Z}_i^{(*)} & = \mathbf{Z}_i^{(t)} + h\, \nabla_\mathbf{Z} \log f(\mathbf{y}_i, \mathbf{Z}_i^{(t)}; \bm\Theta_{\bm\kappa}^{(t)}) + \sqrt{2 h}\, \bm\xi^{(t)} \nonumber\\[2pt]
    & = \mathbf{Z}_i^{(t)} + h\, \left( \nabla_\mathbf{Z} \log f(\mathbf{y}_i \cond \mathbf{Z}_i^{(t)}; \bm\Theta_{\bm\kappa}^{(t)}) + \nabla_\mathbf{Z} \log f(\mathbf{Z}_i^{(t)}; \bm\Theta_{\bm\kappa}^{(t)}) \right)  + \sqrt{2 h}\, \bm\xi^{(t)}\,, \label{eq:SDE_MALA}
\end{align}
where $h \propto K^{1/3}$ is a fixed discretization step, and $\bm\xi^{(t)}$ is a sequence of $K$-dimensional standard Normal random variables.

The discretization of the continuous-time process introduces bias and makes the sampling inexact. To address this issue, the proposal $\mathbf{Z}_i^{(*)}$ in \eqref{eq:SDE_MALA} is accepted or rejected according to the Metropolis-Hastings algorithm, thereby ensuring ergodicity and exact preservation of the invariant measure targeting $p(\mathbf{Z} \cond \mathbf{y}_i; \bm\Theta_{\bm\kappa}^{(t)})$. Namely, for each unit $i = 1,\ldots,N$ in the sample, we set $\mathbf{Z}_i^{(t+1)} = \mathbf{Z}_i^{(*)}$ with probability
\begin{equation}
    \alpha_h(\mathbf{Z}_i^{(*)}, \mathbf{Z}_i^{(t)}) = \min \left\{1, \, \dfrac{f(\mathbf{y}_i, \mathbf{Z}_i^{(*)}; \bm\Theta_{\bm\kappa}^{(t)})\, q_h(\mathbf{Z}_i^{(t)}\, |\, \mathbf{Z}_i^{(*)})}{f(\mathbf{y}_i, \mathbf{Z}_i^{(t)}; \bm\Theta_{\bm\kappa}^{(t)})\, q_h(\mathbf{Z}_i^{(*)}\, |\, \mathbf{Z}_i^{(t)})}\right\}\,, \label{eq:reject_MALA}
\end{equation}
where $q_h(\mathbf{Z}_i'\,|\, \mathbf{Z}_i)$ is the transition kernel given by
\begin{equation}
    q_h(\mathbf{Z}_i'\, |\, \mathbf{Z}_i) = \frac{1}{(4\pi h)^{K/2}}\, \exp \left(-\frac{1}{4h} \big\vert \big\vert \mathbf{Z}_i' - \mathbf{Z}_i - h\, \nabla_\mathbf{Z} \log f(\mathbf{y}_i, \mathbf{Z}; \bm\Theta_{\bm\kappa}^{(t)}) \rvert_{\mathbf{Z} = \mathbf{Z}_i} \, \big\vert \big\vert^2_2\right) \label{eq:kernel_MALA}
\end{equation}

In this paper, we set $h = \mu_z\,K^{-1/3}$ for some constant $\mu_z > 0$ such that the acceptance rates are, on average, between 40\% and 80\% \citep{Roberts&Rosenthal_JRSSB1998}. The convergence of parameter estimates obtained via the MALA-based SA-MD algorithm to MMLE can be established using the theoretical results in \citet{DeBortoliEtAl_Stats&Comp2021}.

Latent attributes in the boundaries (i.e., $\phi(Z_k) = 0$ and $\phi(Z_k) = 1$) are only possible when $Z_k \to -\infty$ and $Z_k \to \infty$, which are measure-zero events under the Gaussian copula parametrization. However, to avoid numerical instabilities caused by limit probabilities of $\pi_j(\mathbf{Z}) = 0$ and/or $\pi_j(\mathbf Z) = 1$ (resulting in $\pi_j(\mathbf{Z})(1-\pi_j(\mathbf{Z})) \approx 0$), we introduce the `clamped' term $\max(\epsilon, \pi_j(\mathbf{Z})(1-\pi_j(\mathbf{Z})))$ for a small $\epsilon > 0$ whenever required (see analytical expressions in Section \ref{sec:OSM_Complexity}). These events, however, are infrequent and are mitigated by the MALA sampler on the $\mathbf{Z}$-space in two ways: first, through the rejection step, by accepting only high probability proposals; and second, through the decaying step size in the SA step, which dampens the influence of large gradient values in later iterations.

\subsection{Mirror Descent (MD) update rules} \label{sec:OSM_MD-rules}

For simplicity, we write $G(\bm\Theta_{\bm \kappa}^{(t-1)})$ as an alias for the approximate stochastic gradient vector $G(\bm\Theta_{\bm \kappa}^{(t-1)}, \mathbf{U}_1^{(t)}, \ldots, \mathbf{U}_N^{(t)})$ in \eqref{eq:stochG}. Moreover, we write $G(\bm\alpha_{j}^{(t-1)})$ to denote the sub-vectors corresponding to the derivatives with respect to weights $\bm\alpha_j$:
\begin{equation*}
G(\bm\alpha_{j}^{(t-1)}) = \sum\limits_{i = 1}^{N} \nabla_{\bm{\alpha}_{j}} \log f(\mathbf{y}_i, \mathbf{U}^{(t)}_{i}; \bm{\Theta}_{\bm\kappa})\lvert_{\bm{\Theta}_{\bm\kappa} = \bm{\Theta}_{\bm\kappa}^{(t-1)}}\,;
\end{equation*}
and a similar notation holds for the vector of sieve approximation parameters $\bm\theta_{jk}$ and the row-vector $\bm l_k$. Analytical expressions are presented in Section \ref{sec:OSM_Complexity}.

The analytical solutions for the projection in the MD update rule in \eqref{eq:Theta[t+1]_SMD} are as follows. For weights $\alpha_{jk} \in \bm\alpha_j$ the update rule has a closed form solution \citep{Kivinen&Warmuth_IC1997, Beck&Teboulle_ORL2003}:
\begin{equation*}
    \alpha_{jk}^{(t+1)} = \dfrac{q_{jk}\,\alpha_{jk}^{(t)}\, \exp(\gamma_\alpha^{(t)}\big[G(\bm\alpha_{j}^{(t)})\big]_k\,)}{\sum_{k'= 1}^{K} q_{jk'}\, \alpha_{jk'}^{(t)}\, \exp(\gamma_\alpha^{(t)}\big[G(\bm\alpha_{j}^{(t)})\big]_{k'}\,)}\,,
\end{equation*}
where $[\bm x]_i$ denotes the $i$-th entry of a vector $\bm x$, $q_{jk} \in \mathbf{q}_j$ is the $k$th entry in the $j$th row of the \textbf{Q}-matrix, and $\gamma_\alpha^{(t)} = \mu_\alpha\, t^{-0.51}$ is the weights-specific step size with constant $\mu_\alpha > 0$. In vector notation, we have:
\begin{equation}
    \bm\alpha_j^{(t+1)} = \frac{\bm\alpha_j^{(t)} \odot \exp(\gamma_\alpha^{(t)}G(\bm\alpha_j^{(t)}))}{\exp(\gamma_\alpha^{(t)}G(\bm\alpha_j^{(t)}))^\top \bm\alpha_j^{(t)}} \odot \mathbf{q}_j\,, \label{eq:mirror_alpha}
\end{equation}
where $\exp(\cdot)$ is element-wise and $\odot$ denotes the entry-wise product.

A similar update rule is derived for sieve parameters $\theta_{jk,l} \in \bm\theta_{jk}$:
\begin{equation*}
    \theta_{jk,l}^{(t+1)} = \dfrac{\theta_{jk,l}^{(t)}\, \exp(\gamma_\theta^{(t)}\big[G(\bm\theta_{jk}^{(t)})\big ]_l\,)}{\sum_{l'= 1}^{L+1} \theta_{jk,l'}^{(t)}\, \exp(\gamma_\theta^{(t)}\big[G(\bm\theta_{jk}^{(t)}\big]_{l'}\,)}\,,
\end{equation*}
with similar expression in vector notation:
\begin{equation}
    \bm\theta_{jk}^{(t+1)} = \frac{\bm\theta_{jk}^{(t)} \odot \exp(\gamma_\theta^{(t)}G(\bm\theta_{jk}^{(t)}))}{\exp(\gamma_\theta^{(t)}G(\bm\theta_{jk}^{(t)}))^\top \bm\theta_{jk}^{(t)}}\,. \label{eq:mirror_theta}
\end{equation}

For non-zero entries in the $k$-th row $\bm l_k$ of the lower triangular matrix $\mathbf{L}$, the MD update rule becomes a projected gradient descent update with closed form solution:
\begin{equation*}
  l_{kk'}^{(t+1)} = \frac{(l_{kk'}^{(t)} + \gamma_l^{(t)}\big[G(\bm l_k^{(t)})\big]_{k'}\,)}{\sum\limits_{k < k'}(l_{kk'}^{(t)} + \gamma_l^{(t)}\big[G(\bm l_k^{(t)})\big]_{k'}\,)^2}\,,
\end{equation*}
or in vector notation:
\begin{equation}
\bm l_k^{(t+1)} = \frac{\bm l_k^{(t)} + \gamma_l^{(t)}G(\bm l_k^{(t)})}{||\bm l_k^{(t)} + \gamma_l^{(t)}G(\bm l_k^{(t)})||^2_2}\,, \label{eq:mirror_L}
\end{equation}

\subsection{Computational complexity analysis} \label{sec:OSM_Complexity}

We give the necessary analytical expressions to study the computational complexity of the proposed SA-MD algorithm. The gradient components of $\nabla_\mathbf{Z} \log f(\mathbf{y}_i, \mathbf{Z}_i; \bm\Theta_{\bm\kappa})$ in the MALA update equation \eqref{eq:SDE_MALA} are:
\begin{align}
    \nabla_\mathbf{Z} \log f(\mathbf{y}_i \cond \mathbf{Z}_i; \bm\Theta_{\bm\kappa}) & = \sum_{j = 1}^J \left[\frac{y_{ij} - \pi_{ij}}{\pi_{ij}\,(1-\pi_{ij})}\right] \nabla_\mathbf{Z}\pi_{ij}\,, \label{eq:complex_fYZ} \\[2pt]
    \nabla_\mathbf{Z} \log f(\mathbf{Z}_i; \bm\Theta_{\bm\kappa}) & = \bm\Sigma^{-1} \mathbf{Z}_i \label{eq:complex_fZ}
\end{align}
with $\nabla_\mathbf{Z}\pi_{ij} = \left(\alpha_{jk}\,\frac{\partial g_{jk}^s(U_{ik})}{\partial U_{k}} \phi(Z_{ik}): k = 1,\ldots,K \right)^\top$.

Let $\pi_{ij} := \pi_j(\mathbf{U}_i;\bm\alpha_j, \bm\theta_j, \bm\kappa)$ and $g_{jk}^s(U_{ik}) := g_{jk}^s(U_{ik}; \bm\theta_{jk}, \bm\kappa)$. For an item $j = 1,\ldots,J$, latent attribute $k = 1,\ldots,K$, and sieve approximation parameter $l = 1,\ldots,L+1$, the entries of the stochastic gradient in equation \eqref{eq:stochG} are of the form:
\begin{align}
    \big[G(\bm\alpha_j)\big]_k & = \sum_{i = 1}^N \left[\frac{y_{ij} - \pi_{ij}}{\pi_{ij}\,(1-\pi_{ij})}\right]  q_{jk}\, g_{jk}^s(U_{ik})\,, \label{eq:complex_alpha} \\[2pt]
    \big[G(\bm\theta_{jk})\big]_l & = \sum_{i = 1}^N \left[\frac{y_{ij} - \pi_{ij}}{\pi_{ij}\,(1-\pi_{ij})}\right] \alpha_{jk} \, q_{jk}\, \frac{U_{ik} - \kappa_{l-1}}{\kappa_{l}-\kappa_{l-1}}\,, \label{eq:complex_theta}
\end{align}
with the convention $\kappa_0 = 0$ and $\kappa_{L+1} = 1$ for boundary knots; and for $k$-th row $\bm l_k$ in $\mathbf{L}$:
\begin{equation}
    \big[G(\bm{l}_k)\big]_{k'} = - N \, \text{tr}\left( \mathbf{L}^\top (\mathbf{L}\mathbf{L}^\top)^{-1}\mathbf{D}_{kk'} \right) +  \sum\limits_{i=1}^{N} \mathbf{Z}_i^\top \mathbf{G}_{kk'} \mathbf{Z}_i \,, \label{eq:complex_L}
\end{equation}
where $k'$ iterates over non-zero entries of $\bm l_k$, $\mathbf{D}_{kk'} = \partial \mathbf{L} / \partial l_{kk'}$ is a square matrix of dimension $K$ with a value of 1 in the $(k,k')$-th position and zero elsewhere, and $\mathbf{G}_{kk'} = (\mathbf{LL}^\top)^{-1} \mathbf{D}_{kk'} \mathbf{L}^\top (\mathbf{L} \mathbf{L}^\top)^{-1}$. For computational simplicity and given the structure of $\mathbf{D}_{kk'}$, we reparametrize $\mathbf{Z}_i^\top \mathbf{G}_{kk'} \mathbf{Z}_i = [\mathbf{W}_i]_k \times [\mathbf{V}_i]_{k'}$, with $K$-dimensional vectors $\mathbf{W}_i = (\mathbf{LL}^\top)^{-1}\mathbf{Z}_i$ and $\mathbf{V}_i = \mathbf{L}^\top\mathbf{W}_i$. 

The per-iteration cost of the proposed SA-MD algorithm sketched in Algorithm \ref{alg:SA-MD} is computed as follows. The MALA sampling step in the SA-step \eqref{eq:poster} requires evaluating:
\begin{enumerate}
    \item[M1.] \textbf{IRF evaluation}
    \begin{itemize}
        \item[a.] $U_{ik} = \phi(Z_{ik})$; $k = 1,\dots,K$ \hfill $O(K)$
        \item[b.] Assigning $U_{ik}$ to interval in grid defined by $\bm\kappa$; $k = 1,\ldots,K$ \hfill $O(K)$
        \item[c.] $g^s_{jk}(U_{ik})$; $j = 1,\ldots,J$, $k = 1,\ldots,K$ \hfill $O(JK)$
        \item[d.] $\pi_{ij} = \sum_{k} \alpha_{ij}\,g^s_{jk}(U_{ik})$; $j = 1,\ldots,J$, $k = 1,\ldots,K$ \hfill $O(JK)$ 
        \item[] \textbf{Total cost} M1 (over $i = 1,\ldots,N$) \hfill $O(NJK)$
    \end{itemize}
    \item[M2.] \textbf{Gradient evaluation} \qquad (\ref{eq:complex_fYZ} and \ref{eq:complex_fZ})
    \begin{itemize}
        \item[a.] $\dfrac{y_{ij}-\pi_{ij}}{(\pi_{ij}(1-\pi_{ij}))}$ (precomputed $\pi_{ij}$s from M1); $j = 1,\ldots,J$ \hfill $O(J)$
        \item[b.] $\nabla_{\mathbf{Z}}\pi_{ij}$ (known positions of $\mathbf{U}_{i}$ in grid from M1b) \hfill $O(K)$
        \item[c.] $\nabla_{\mathbf{Z}}\log f(\mathbf{y}_i \cond \mathbf{Z}_i;\bm\Theta_{\bm\kappa})$ (\ref{eq:complex_fYZ}, from M2a and M2b) \hfill $O(JK)$
        \item[d.] $\nabla_{\mathbf{Z}}\log f(\mathbf{Z}_i;\bm\Theta_{\bm\kappa})$ (\ref{eq:complex_fZ}, $\bm\Sigma^{-1}$ precomputed, else $O(K^3)$) \hfill $O(K^2)$
        \item[]  \textbf{Total cost} M2 (over $i = 1,\ldots,N$) \hfill $O(NJK + NK^2)$
    \end{itemize}
    \item[M3.] \textbf{Sampling and rejection steps} \qquad (\ref{eq:SDE_MALA} and \ref{eq:reject_MALA})
    \begin{itemize}
        \item[a.] $\mathbf{Z}_i^{(*)}$ (\ref{eq:SDE_MALA}, with precomputed gradients from M2c and M2d) \hfill $O(K)$
        \item[b.] $\alpha_h(\mathbf{Z}_i^{(*)}, \mathbf{Z}_i^{(t)})$ (\ref{eq:reject_MALA}, from computing \eqref{eq:kernel_MALA} again) \hfill $O(JK + K^2)$
        \item[] \textbf{Total cost} M3 (over $i = 1,\ldots,N$) \hfill $O(NJK + NK^2)$
    \end{itemize}
    \item[] \textbf{Total cost} M1 + M2 + M3 \hfill $O(NJK + NK^2)$
\end{enumerate}

With new $\mathbf{U}_i$, $i=1,\ldots,N$ available, computing the stochastic gradient in \eqref{eq:stochG} requires evaluating:
\begin{enumerate}
    \item[G1.] \textbf{IRF evaluation and components}\hfill $O(JK)$ 
    \begin{itemize}
        \item[a.] $\pi_{ij}$; $j = 1,\ldots,J$, $k = 1,\ldots,K$ (same as M1 above) \hfill $O(JK)$ 
        \item[b.] $\dfrac{y_{ij}-\pi_{ij}}{(\pi_{ij}(1-\pi_{ij}))}$ (same as M2a above); $j = 1,\ldots,J$ \hfill $O(J)$
        \item[] \textbf{Total cost} G1 (over $i = 1,\ldots,N$) \hfill $O(NJK)$
    \end{itemize}
    \item[G2.] \textbf{Gradient evaluation} \qquad (\ref{eq:complex_alpha} to \ref{eq:complex_L})
    \begin{itemize}
        \item[a.] $G(\bm\alpha_j)$ (from G1 and \ref{eq:complex_alpha}); $j = 1,\ldots,J$ \hfill $O(NJK)$
        \item[b.] $G(\bm\theta_{jk})$ (from G1 and \ref{eq:complex_theta}); $j = 1,\ldots,J$, $k = 1,\ldots,K$ \hfill $O(NJKL)$
        \item[c.] $G(\bm{l}_k)$ \eqref{eq:complex_L}
        \begin{itemize}
            \item[I.] $N \, \text{tr}\left( \mathbf{L}^\top (\mathbf{L}\mathbf{L}^\top)^{-1}\mathbf{D}_{kk'} \right)$ \hfill $O(K^3)$
            \item[II.] $\mathbf{Z}_i^\top \mathbf{G}_{kk'} \mathbf{Z}_i = [\mathbf{W}_i]_k \times [\mathbf{V}_i]_{k'}$; $i = 1,\ldots,N$ \hfill $O(NK^2)$
            \item[] Total cost G2c \hfill $O(NK^2 + K^3)$
        \end{itemize}
        \item[]  \textbf{Total cost} G2 ($N \gg K$) \hfill $O(NJKL + NK^2)$
    \end{itemize}
    \item[] \textbf{Total cost} G1 + G2 \hfill $O(NJKL + NK^2)$
\end{enumerate}

With the stochastic gradient $G(\bm\Theta_{\bm\kappa})$ available, computing the MD-step \eqref{eq:Theta[t+1]_SMD} using update equations \eqref{eq:mirror_alpha} to \eqref{eq:mirror_L} requires evaluating:
\begin{enumerate}
    \item[D1.] Exponentiated gradient for $\bm\alpha_j$ \eqref{eq:mirror_alpha}; $j = 1,\ldots,J$ \hfill $O(JK)$
    \item[D2.] Exponentiated gradient for $\bm\theta_{jk}$ \eqref{eq:mirror_theta}; $j = 1,\ldots,J$, $k = 1,\ldots,K$ \hfill $O(JKL)$
    \item[D3.] Projected gradient for $\bm l_k$ (\ref{eq:mirror_L}, by normalization) ; $k = 1,\ldots,K$ \hfill $O(K^2)$
    \item[] \textbf{Total cost} D1 + D2 + D3 \hfill $O(JKL + K^2)$
\end{enumerate}

Thus, the total cost per-iteration of the SA-MD algorithm in \ref{alg:SA-MD} is $O(NJKL + NK^2)$. Since $JL \gg K$ in most settings, the per-iteration cost becomes $O(NJKL)$.

\section{Computing the marginal log-likelihood via Importance Sampling}  \label{sec:OSM_mllk+IS}

The marginal log-likelihood is a central quantity used for model comparison and assessing goodness-of-fit. In general, marginal (log-)likelihoods are analytically intractable and difficult to compute for most LVMs, especially in models with a large number of latent attributes. In this paper, we compute the marginal log-likelihood using a Monte Carlo integration approach via importance sampling (IS).

The IS approximation of the marginal likelihood contribution for an individual $i$ evaluated at the SMMLE $\hat{\bm{\Theta}_{\bm\kappa}}$ is:
\begin{equation*}
    \int_{\R^K} f(\mathbf{y}_i \cond \mathbf{Z}; \hat{\bm{\Theta}}_{\bm\kappa})\, p(\mathbf{Z}; \hat{\bm{\Theta}}_{\bm\kappa})\,  \text{d}\mathbf{Z} \approx \frac{1}{M} \sum_{m = 1}^M \frac{f(\mathbf{y}_i \cond \mathbf{z}_m; \hat{\bm{\Theta}}_{\bm\kappa})\, p(\mathbf{z}_m; \hat{\bm{\Theta}}_{\bm\kappa})}{q_i(\mathbf{z}_m)}\,; \qquad \mathbf{z}_m \sim q_i(\mathbf{Z})\,,
\end{equation*}
where $q_i(\mathbf{Z})$ is the \textit{importance distribution} (indexed by $i$) and $M$ is the number of IS iterations. The marginal likelihood computed via IS will have lower variance than the naive Monte-Carlo estimate for an appropriately chosen $q_i(\mathbf{Z})$. This can be achieved by sampling from an importance distribution with heavier tails than $p(\mathbf{Z};\hat{\bm{\Theta}}_{\bm\kappa})$. For convenience, we choose $q_i(\mathbf{Z}) = \text{MVN}(\bm{\hat{\mu}}_i, \hat{\bm{\Sigma}}_i + \mathbb{I}_K)$, where $\hat{\bm{\mu}}_i$ and $\hat{\bm{\Sigma}}_i$ are the Polyak-Ruppert trajectory averages of the posterior mean and posterior covariance matrix of the latent attributes for unit $i$, computed via the online estimators $\bm{\mu}_i^{(t+1)} = \bm{\mu}_i^{(t)} + (\mathbf{z}_i^{(t)} - \bm{\mu}_i^{(t)})/(t+1)$ for the mean, and $\bm{\Sigma}_i^{(t+1)} = (t\,\bm{\Sigma}_i^{(t)})/(t+1) + (t\,(\mathbf{z}_i^{(t)} - \bm{\mu}_i^{(t)})(\mathbf{z}_i^{(t)} - \bm{\mu}_i^{(t)})^\top)/(t+1)^2$ for the covariance matrix, respectively (see, \citealp{Dasgupta&Hsu_COLT2007}).

\section{Additional material on simulation studies} \label{sec:OSM_Simulation}

We now describe details and implementation settings for the confirmatory and exploratory simulation studies. In both settings, we consider following \textbf{Q}-matrices:
\begin{equation*}
\mathbf{Q}_3^\top = \begin{bmatrix}
1 & 0 & 0 & 1 & 0 & 0 & 1 & 0 & 0 & 1 & 1 & 0 & 1 & 1 & 0 & 1 & 1 & 1 & 1 & 1 \\
0 & 1 & 0 & 0 & 1 & 0 & 0 & 1 & 0 & 1 & 0 & 1 & 1 & 0 & 1 & 1 & 0 & 1 & 1 & 1 \\
0 & 0 & 1 & 0 & 0 & 1 & 0 & 0 & 1 & 0 & 1 & 1 & 0 & 1 & 1 & 0 & 1 & 1 & 1 & 1 \\
\end{bmatrix}
\label{eq:App_S1_Q3}
\end{equation*}

\begin{equation*}
\mathbf{Q}_5^\top = \begin{bmatrix}
1 & 0 & 0 & 0 & 0 & 1 & 0 & 0 & 0 & 0 & 1 & 0 & 0 & 0 & 1 & 1 & 0 & 0 & 1 & 1 \\
0 & 1 & 0 & 0 & 0 & 0 & 1 & 0 & 0 & 0 & 1 & 1 & 0 & 0 & 0 & 1 & 1 & 0 & 0 & 1 \\
0 & 0 & 1 & 0 & 0 & 0 & 0 & 1 & 0 & 0 & 0 & 1 & 1 & 0 & 0 & 1 & 1 & 1 & 0 & 0 \\
0 & 0 & 0 & 1 & 0 & 0 & 0 & 0 & 1 & 0 & 0 & 0 & 1 & 1 & 0 & 0 & 1 & 1 & 1 & 0 \\
0 & 0 & 0 & 0 & 1 & 0 & 0 & 0 & 0 & 1 & 0 & 0 & 0 & 1 & 1 & 0 & 0 & 1 & 1 & 1 \\
\end{bmatrix}
\label{eq:App_S1_Q5}
\end{equation*}

For the IRFs in the aPM-CDM, the true guessing probabilities (intercepts) are drawn from $\delta_{j0} \sim \mathrm{Unif}(0, 0.2)$ for all $j = 1,\ldots,J$. The slipping probabilities, denoted by $s_j$, is also drawn from $s_j \sim \mathrm{Unif}(0,0.2)$ for all items. The non-zero factor loadings in the aPM-CDM $\{\delta_{jk}: q_{jk} \neq 0 \,,\,k = 1,\ldots,K\}$ are  generated from $\delta_{jk} \sim \mathrm{Unif}(0,1-\delta_{j0}-s_j)$ and then adjusted such that $1-||\bm{\delta}_j||_1 = s_j$. 

The starting values for the intercepts (guessing probability) are $\delta_{j0}^{(0)} = 0.1$, but can be adjusted by the control arguments in the \texttt{apmCDM} function in the accompanying \texttt{R} package \texttt{gapmCDM}. The starting values for the non-zero factor slopes are set to equal values such that $1 - \sum_{k=0}^K \delta_{jk}^{(0)} = 0.1$, a starting user-defined slipping probability, also subject to adjustment. For example, when $K = 3$, the first row of the \textbf{Q}$_3$-matrix is $\mathbf{q}_1 = (1,0,0)^\top$ and thus the initial values for the aPM-CDM IRF for item 1 are $\bm{\delta}_1^{(0)} = (0.1,0.8,0,0)^\top$. For item 10, we have $\mathbf{q}_{10} = (1,1,0)^\top$, and therefore $\bm{\delta}_{10}^{(0)} = (0.1,0.4,0.4,0)^\top$. Likewise, for item 20, the corresponding row is $\mathbf{q}_{20} = \bm{1}_3^\top$ and thus the IRF for this item is parametrized by $\bm{\theta}_{20} = (0.1,0.3,0.3,0.3)^\top$. We follow a similar strategy when $K = 5$.

For the IRFs in the GaPM-CDM, the true weights are set to equal values such that $\mathbf{q}_j^\top \bm\alpha_j = 1$. In the confirmatory setting, we follow the same approach. For example, when $K = 3$, the first row of the \textbf{Q}$_3$-matrix is $\mathbf{q}_1 = (1,0,0)^\top$ and thus the true and initial values for the weights in the GaPM-CDM IRF for item 1 are $\bm{\alpha}_1^{(0)} = (1,0,0)^\top$. For item 10, we have $\mathbf{q}_{10} = (1,1,0)^\top$, and thus $\bm{\alpha}_{10}^{(0)} = (0.5,0.5,0)^\top$, etc. In the exploratory setting, the starting weights are set to equal values such that $||\bm\alpha_j||_1 = 1$ for all items. The same approach holds for $K = 5$. For both confirmatory and exploratory GaPM-CDMs, the starting values for the sieve approximation parameters are set to values such that $g^{s}_{jk}(U_k; \bm{\theta}_{jk}^{(0)}, \bm{\kappa})$ is the identity function defined on $[0,1]$, for all $j = 1,\ldots,J$ and $k = 1,\ldots,K$.

For both models and settings, the initial vector of means for the latent attributes is set to $\bm{\mu}_K^{(0)} = \bm{0}_K$ and the covariance matrix to $\bm{\Sigma}_K^{(0)} = \mathbb{I}_K$, with $K \in \{3,5\}$. Recall that, for model identification in the GaPM-CDM, $\bm{\mu}_K = \bm{0}_K$ is fixed, and $\bm{\Sigma}_K$ is constrained to be a correlation matrix. The starting values for the latent attributes, $(\mathbf{Z}_i^{(0)}: i = 1,\ldots,N)$ are the regression-based factor scores from an exploratory factor analysis on the tetrachoric correlation of the observed (binary) data, after performing the oblimin rotation.

When estimating both GaPM-CDM and PM-ACDM models, we set the number of iterations for the SA-MD algorithm to $T = 90000$, with a burn-in period of $\omega = 45000$. The constants in the decaying step size (see Section \ref{sec:OSM_SA-MD}) were fine-tuned for different types of parameters under different settings. We set $\mu_{\delta}$ and $\mu_l$ to $1/N$ for the confirmatory aPM-CDM when the true model was the aPM-CDM, but set to $\mu_{\delta} = \mu_l = 0.3/N$ when the true model was the GaPM-CDM. For the GaPM-CDM, we fixed $\mu_\alpha$, $\mu_\theta$, and $\mu_l$ to $1/N$ in in both the confirmatory and exploratory settings. The discretization step in the MALA sampler is set to $h = 0.2$ if $K = 3$, and $h = 0.1$ if $K = 5$. This results in average acceptance rates between 60\% and 90\% on the training sample across all models and simulation settings. 

\section{Additional material on empirical applications} \label{sec:OSM_Apps}

\subsection{English test data example} \label{sec:OSM_Apps_ECPE}

We follow a similar strategy as in the simulation studies when defining the starting values for the parameters and latent attributes in the ECPE dataset. Similarly, we run the SA-MD algorithm for $T = 90000$ iterations, with a burn-in period of $\omega = 45000$. The discretization step is set to $h = 0.1$ for both the aPM-CDM and the GaPM-CDM, which leads to average acceptance rates between 70\% and 85\% on the training sample in the cross-validation exercise across both models.

For completeness and model comparison, we present the results for the computed deviance (defined as $-2 \ell_N(\hat{\bm\Theta}_{\bm\kappa})$ for the GaPM-CDM, and $-2 \ell_N(\hat{\bm\Theta})$ for the aPM-CDM, ACDM, and GDINA). We see how the GaPM-CDM fits the ECPE data better than the competing aPM-CDM and traditional CDMs.

\begin{table}[!ht]
\vspace{0.3cm}
\begin{tabular}{ccccc} \toprule
Model & GaPM-CDM ($\bm\kappa_1$) & aPM-CDM & ACDM & GDINA \\ \midrule
$-2\ell_N(\hat{\bm\Theta})$ & \textbf{84866.84} & 85037.85 & 85491.10 & 85479.54 \\
\bottomrule
\end{tabular}
\caption{Deviance results for the ECPE dataset.}
\label{tab:ECPE_fit}
\end{table}

\begin{figure}[!ht]
\vspace{5mm}
    \centering
    \begin{subfigure}[t]{\textwidth}
        \centering
        \includegraphics[scale = 0.45]{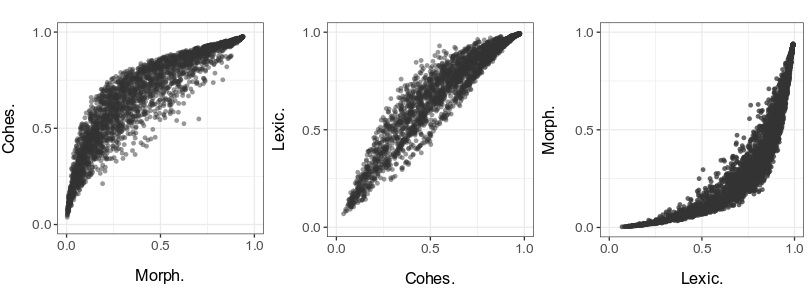}
        \caption{EAP factor scores for aPM-CDM}
        \label{fig:OSM_ECPE_EAP_aPMCDM}
    \end{subfigure}
    \\
    \begin{subfigure}[t]{\textwidth}
        \centering
        \includegraphics[scale = 0.45]{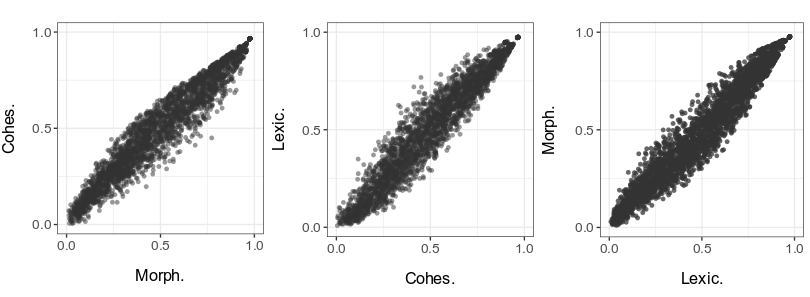}
        \caption{EAP factor scores for GaPM-CDM}
        \label{fig:OSM_ECPE_EAP_GaPMCDM}
    \end{subfigure}
      \caption{Estimated EAP factor scores. ECPE data.}
    \label{fig:OSM_ECPE_EAP}
\end{figure}

Figure \ref{fig:OSM_ECPE_EAP} shows the estimated EAP factor scores for both models. The scatterplots for the aPM-CDM factor scores (Figure \ref{fig:OSM_ECPE_EAP_aPMCDM}) are consistent with those in \citet{ShangEtAl_AoAS2021}. In the GaPM-CDM (Figure \ref{fig:OSM_ECPE_EAP_GaPMCDM}), we are constrained by the fixed means in the Gaussian copula. However, the factor scores produced by the two models are largely consistent in terms of their rank order. In fact, the Spearman's rank correlations between factor scores are $0.99$ for all three attributes.

\subsection{PROMIS example} \label{sec:OSM_Apps_PROMIS}

We use the same initialization strategy for defining the starting values for the parameters and latent attributes in the PROMIS dataset. We run the SA-MD algorithm for $T = 90000$ iterations, with a burn-in period of $\omega = 45000$. The discretization step is set to $h = 0.1$ for both the aPM-CDM and the GaPM-CDM across all potential values for the latent attributes, which leads to average acceptance rates between 15\% and 70\% on the training sample in the cross-validation exercise for both models and across $K = 1,\ldots,7$. The acceptance rate for $K = 1$ is around 30\% in both cases, peaks at around 70\% when $K = 5$, and decreases rapidly to 15\% when $K = 7$.

For the GaPM-CDM, the constants in the decaying step size are set to $\mu_\alpha = \mu_\theta = 1/N$, and $\mu_l = 0.75/N$ for $K < 5$ and $\mu_l = 0.5/N$ when $K \geq 5$. For the aPM-CDM, we set $\mu_\delta = 0.3/N$, and $\mu_l = 0.20/N$ for $K < 5$ and $\mu_l = 0.1/N$ when $K \geq 5$. These values are chosen for computational stability based on extensive experimentation.

\begin{figure}[!ht]
\centering
\vspace{5mm}
\includegraphics[scale = 0.65]{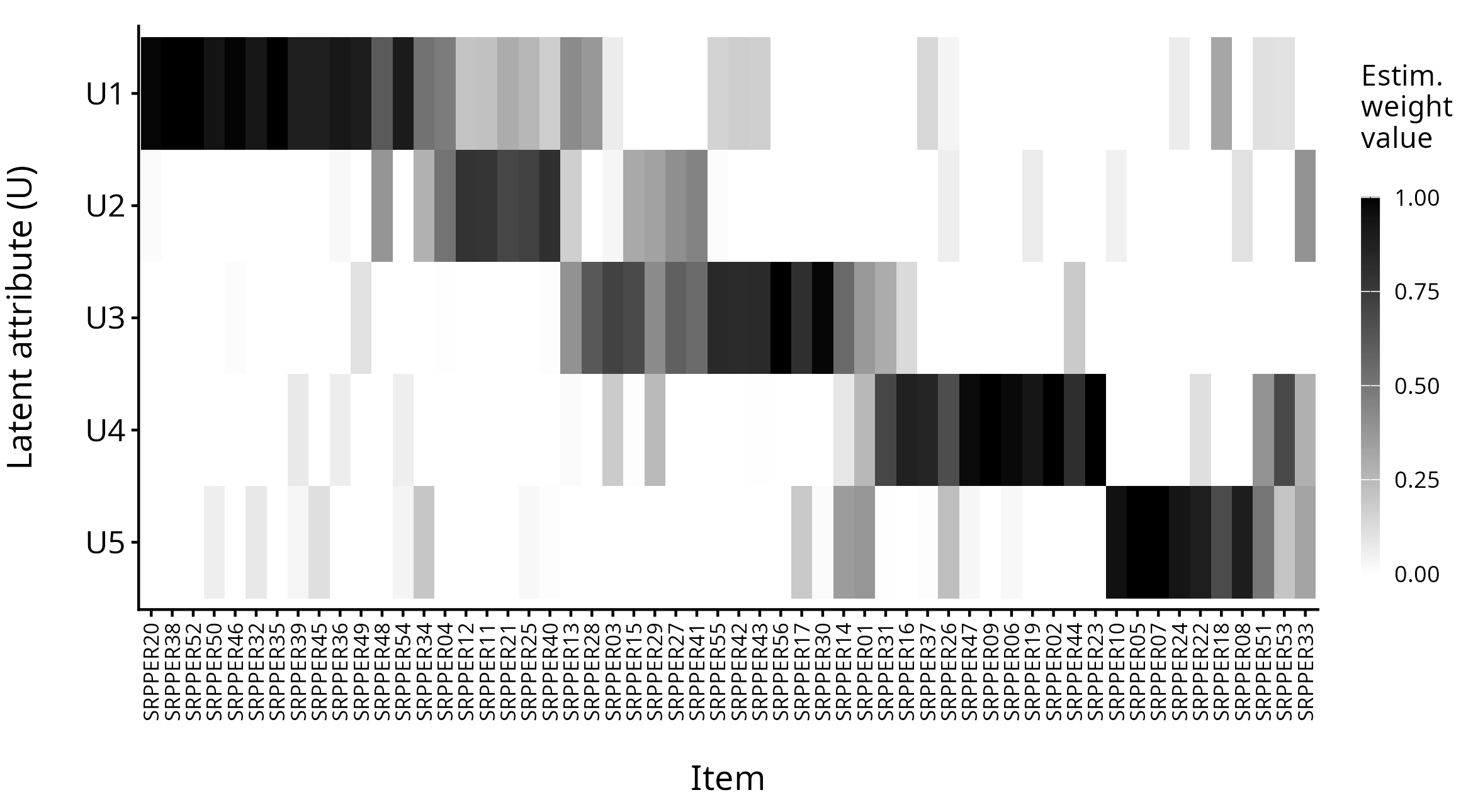}
\caption{Matrix of estimated factor loadings (transposed) for the $\hat K = 5$ dimensional GaPM-CDM on the PROMIS dataset, with $L = 6$ equally spaced knots.}
\label{fig:OSM_PROMIS_Ahat_newK}
\end{figure}

We include a sensitivity analysis to the number of knots $L$. Following the identification results discussed in Remark 4, we fit the GaPM-CDM with a number of knots equal to the lower bound given by $L = \lceil\, (J-1)/2K\, \rceil = 6$. The matrix of estimated weights in Figure \ref{fig:OSM_PROMIS_Ahat_newK} keeps an interpretable and sparse solution, similar to that in Figure 4 in the main text. As expected, the deviance in Table \ref{tab:PROMIS_fit} suggests that the model with a lower number of knots fits slightly worse than the one with 19 knots.

\begin{figure}[!ht]
\centering
\vspace{5mm}
\includegraphics[scale = 0.55]{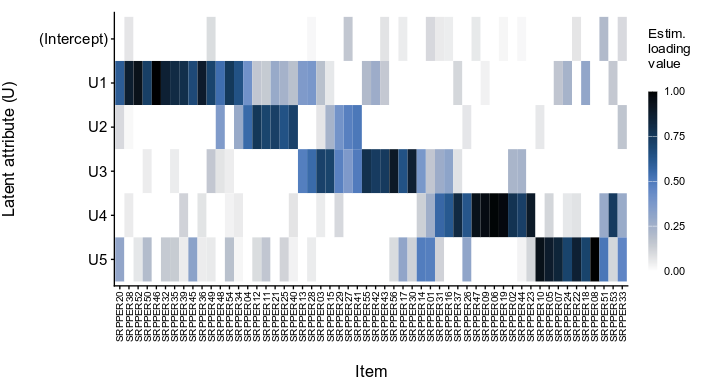}
\caption{Matrix of estimated factor loadings (transposed) for the $K = 5$ dimensional aPM-CDM on the PROMIS dataset.}
\label{fig:A2_PROMIS_Ghat}
\end{figure}

For completeness, we also include the matrix of intercepts and factor loadings from the exploratory aPM-CDM with $\hat K = 5$ (Figure \ref{fig:A2_PROMIS_Ghat}). Note how the intercept term is close to zero in most cases, suggesting low-to-none guessing behavior in this example. The structure is sparse and interpretable, but the model fit is considerably worse than the GaPM-CDM (Table \ref{tab:PROMIS_fit}).

\begin{table}[!ht]
\vspace{0.3cm}
\begin{tabular}{cccc} \toprule
Model & GaPM-CDM (full knots) & GaPM-CDM (min. knots )& aPM-CDM \\ \midrule
$-2\ell_N(\hat{\bm\Theta})$ & \textbf{26564.60} & 26658.41 & 27434.34 \\
\bottomrule
\end{tabular}
\caption{Deviance results for the PROMIS dataset.}
\label{tab:PROMIS_fit}
\end{table}

\section{Theoretical results} \label{sec:OSM_proof}

This section proves Proposition 1 in the paper. We first show in Section \ref{sec:OSM_RLCM} that the isotonic model in Remark 4 can be represented by an equivalent restricted latent class model (RLCM). We then introduce the notion of generic identifiability in Section \ref{sec:OSM_GI}. Section \ref{sec:OSM_proofS1} presents the proof of Proposition 1. Finally, Section \ref{sec:OSM_approx} presents the formal result and proof for approximating \texorpdfstring{$g_{jk}^s$}{} by \texorpdfstring{$g_{jk}^{\text{step}}$}{}.

\subsection{RLCM representation of GaPM-CDM with piecewise constant \texorpdfstring{$g_{jk}^s$}{}}\label{sec:OSM_RLCM}

As discussed in Remark 4, a key step in the proof of the proposition is to represent the model in which each $g_{jk}^s$ is restricted to the class of step (piecewise constant) functions, denoted by $g_{jk}^{\text{step}}$. Under this parametrization, the sieve approximation can be written as:
\begin{equation}
    g_{jk}^{\text{step}}(x; \bm{s}_{jk}, \bm\kappa) = \sum_{l=1}^{L+1} s_{jk,l}\, b_l(x)\,, \label{eq:OSM_step}
\end{equation}
where $b_l(x) = \bm{1}(x \in [\kappa_{l-1}, \kappa_{l}))$, $l = 1,\ldots,L$, are indicator functions evaluating the condition $x \in [\kappa_{l-1}, \kappa_{l})$; $b_{L+1}(x) = \bm{1}(x \in [\kappa_{L}, 1])$; and $\bm s_{jk} = (s_{jk,l}: l = 2,\ldots,L)^\top$ with $0 = s_{jk,1} \leq \dots \leq s_{jk,L+1} = 1$ when $q_{jk} = 1$. In the sequel, let $\mathcal{A} = \{\alpha_{jk}: q_{jk}=1;\, j = 1,\dots,J\}$ and $\mathcal S = \{\bm s_{jk}: q_{jk}=1;\, j= 1, \dots, J;\, k = 1,\dots, K\}$ be model parameters under the step function approximation. The IRFs become
\begin{equation*}
 \pi_j(\mathbf{U}) = \sum_{k = 1}^{K} \alpha_{jk}\,q_{jk} \sum_{l=1}^{L+1}s_{jk,l}\,b_l(U_k)\,,\qquad j = 1,\ldots,J   
\end{equation*}

We now show that the GaPM-CDM with IRFs defined by piecewise constant functions admits an equivalent RLCM representation.

Let $\bzeta = (\zeta_k: k = 1,\ldots,K)^{\top} \in \{1, \dots, L+1\}^{K}$ be a vector of latent classes. The probability distribution of $\bzeta$ conditional on $\bU$ is $p_{\bzeta \cond \bU} = \prod_{k=1}^{K} \left(\prod_{l=1}^{L+1}b_{l}(U_k)^{\bm 1(\zeta_{k} = l)}\right)$. It can be verified that
\begin{equation*}
\pi_j(\bU) = \sum_{\bzeta \in \{1, \dots, L+1\}^{K}} \eta_{j,\bzeta}~ p_{\bzeta \cond \bU},
\end{equation*}
where
\begin{align}\label{eq:OSM_etajzeta}
    \eta_{j, \bzeta} = \sum_{k=1}^{K} \alpha_{jk}\, q_{jk}\, s_{jk,\zeta_k}.
\end{align}

Note that the indicator functions $b_l(U_k)$ induce a partition of $[0,1]^K$ into $(L+1)^K$ rectangular cells, one for each $\bzeta \in \{1,\ldots,L+1\}^{K}$. Thus, $\pi_j(\bU) = \eta_{j,\bzeta}$ on a cell $\bzeta$ where each $U_k$ falls in the interval $\zeta_k$ (i.e., when $b_l(U_k) = \bm 1(\zeta_k = l)$ for all $k$ and $l$). Additionally, the marginal probability mass function of $\bzeta$ is $\pi_{\bzeta} = \int_{[0,1]^{K}} p_{\bzeta \mid \bU}~\text{d} D(\bU;\bm{\Sigma}) = \int_{\bzeta} \text{d}D(\bU;\bm\Sigma)$. Let $\mathcal{H} = (\eta_{j, \bzeta})_{J \times (L+1)^{K}}$ denote the matrix whose rows are indexed by $j = 1,\ldots,J$ and columns by $\bzeta$, and define $\bm{\pi}_{\bzeta} = (\pi_{\bzeta}: \bzeta \in \{1, \dots, L+1\}^{K})^\top$.

Let $\yy = (y_j : j =1,\ldots,J)^\top$ be a realization of $\bY$, $P_{\text{RLCM}}(\bY = \yy ;\mathcal{H}, \bm{\pi}_{\bzeta})$ denote the probability mass function of $\yy$ under the RLCM representation, and $P_{\text{step}}(\bY = \yy ; \mathcal A, \mathcal S , \bm\Sigma)$ the probability mass function under the GaPM-CDM with IRFs approximated by step-functions as described above. The equivalence of the two representations is established by verifying that 
\begin{align}\label{eq:OSM_equiv_RLCM-step}
    P_{\text{step}}(\bY = \yy ;\, \mathcal A, \mathcal S, \bm{\Sigma}) & = \int_{[0,1]^{K}} \prod_{j=1}^{J} \pi_j(\bU)^{y_j}(1 - \pi_j(\bU))^{1-y_j}\, \text{d} D(\bU;\bm{\Sigma}) \nonumber\\[2pt]
    & = \sum_{\bzeta \in \{1, \dots, L+1\}^{K}} \int_{\bzeta}\, \prod_{j=1}^{J}\eta_{j, \bzeta}^{y_j}(1-\eta_{j, \bzeta})^{1-y_j} \, \text{d} D(\bU;\bm{\Sigma}) \nonumber \\[2pt]
    & = \sum_{\bzeta \in \{1, \dots, L+1\}^{K}} \pi_{\bzeta} \prod_{j=1}^{J}\eta_{j, \bzeta}^{y_j}(1-\eta_{j, \bzeta})^{1-y_j} \nonumber \\[2pt]
    & = P_{\text{RLCM}}(\bY = \yy ;\, \mathcal{H}, \bm{\pi}_{\bzeta})\,
\end{align}
where the step from the first to the second line follows from $\pi_j(\bU) = \eta_{j,\bzeta}$ being constant in cell $\bzeta$, and the step from the second to the third line from the fact that $\prod_{j=1}^{J}\eta_{j, \bzeta}^{y_j}(1-\eta_{j, \bzeta})^{1-y_j}$ does not depend on $\bU$.

\subsection{Generic identifiability}\label{sec:OSM_GI}
We now define the notion of generic identifiability for the RLCM representation. Let $\mathcal{T}$ denote the parameter space of $(\mathcal{H}, \bm{\pi}_{\bzeta})$. We say that $(\mathcal{H}, \bm{\pi}_{\bzeta})$ is strictly identifiable on $\mathcal{T}$, if for any $(\mathcal{H}, \bm{\pi}_{\bzeta})\in \mathcal{T}$, there is no  $(\bar{\mathcal{H}}, \bar{\bm{\pi}}_{\bzeta}) \neq (\mathcal{H}, \bm{\pi}_{\bzeta})$ on $\mathcal{T}$ such that 
\begin{align*}
    P_{\text{RLCM}}(\bY = \yy ;\, \mathcal{H}, \bm{\pi}_{\bzeta}) = P_{\text{RLCM}}(\bY = \yy ;\, \bar{\mathcal{H}}, \bar{\bm{\pi}}_{\bzeta}), \!\quad \text{ for all } \yy \in \{0,1\}^{J}.
\end{align*}

Generic identifiability is closely related to the concept of algebraic variety in algebraic geometry. In particular, an algebraic variety is defined as the simultaneous zero-set of a finite collection of multivariate polynomials $\{f_i\}_{i=1}^{n} \subseteq \mathbb{R}[x_1, \dots, x_d]$, $\mathcal{V} = \mathcal{V}(f_1, \dots, f_n) = \{\mathbf{x}\in \mathbb{R}^{d} \mid f_i(\mathbf{x}) = 0, 1 \leq i \leq n\}$ \citep{AllmanEtAl_AoS2009, Gu&Xu_AoS2020}. An algebraic subvariety $\mathcal{V}$ is all of $\mathbb{R}^{d}$ only when all the polynomials defining it are zero polynomials; otherwise, $\mathcal{V}$ is called a proper subvariety and is of dimension less than $d$, hence necessarily of Lebesgue measure zero in $\mathbb{R}^{d}$. The same argument holds when $\mathbb{R}^{d}$ is replaced by the parameter space $\mathcal{T} \subseteq \mathbb{R}^{d}$ that has full dimension in $\mathbb{R}^{d}$. We are now ready to present the following definition of generic identifiability for the RLCM representation. The generic identifiability for the model described in Remark 4 is defined similarly.

\begin{definition}[Generic identifiability]
The RLCM is said to be generically identifiable on the parameter space $\mathcal{T}$, if $(\mathcal{H}, \bm{\pi}_{\bzeta})$ are strictly identifiable on  $\mathcal{T} ~\backslash ~ \mathcal{V}$ , where $\mathcal{V}$ is a proper algebraic subvariety of $\mathcal{T}$.
\end{definition}

\subsection{Proof of Proposition 1 in the paper}\label{sec:OSM_proofS1}

We give a formal statement for Proposition 1 in the paper:
\setcounter{proposition}{-1}
\begin{proposition}[Proposition 1 in paper] \label{prop:OSM_identif}
In confirmatory settings, if the \textbf{Q}-matrix satisfies Condition 1, then $\mathcal A$ and $\mathcal S$ are generically identified. In exploratory settings, the sufficient condition reduces to $J \geq 2KL + 1$.
\end{proposition}

The following proposition is key to establishing Proposition \ref{prop:OSM_identif}.

\begin{proposition}\label{prop:OSM_ident_asRLCM}
If the \textbf{Q}-matrix satisfies Condition 1, then $(\mathcal{H}, \bm{\pi}_{\bzeta})$ are generically identifiable, up to label swapping among those latent classes that have identical column vectors in $\mathcal{H}$.
\end{proposition}

Before proving the Proposition \ref{prop:OSM_ident_asRLCM}, we first show how it implies Proposition \ref{prop:OSM_identif}. Since Proposition \ref{prop:OSM_identif} concerns only the identifiability of $(\mathcal{A},\mathcal{S})$, we treat $\bm{\Sigma}$ as known. Moreover, the argument in the proof of Proposition \ref{prop:OSM_ident_asRLCM} continues to hold in the special case where $\bm{\pi}_{\bzeta}$ is known. 

By definition, the entries of $\mathcal{H} = (\eta_{j,\bzeta})_{J \times (L+1)^K}$ are polynomial functions of $(\mathcal{A},\mathcal{S})$. Hence, the pre-image of $\mathcal{V}$ under this mapping is an algebraic subvariety. As will be shown in the proof of Proposition \ref{prop:OSM_ident_asRLCM}, there exists a point in $\mathcal{T} \setminus \mathcal{V}$ constructed from a specific choice of $(\mathcal{A},\mathcal{S})$; therefore, the pre-image of $\mathcal{V}$ is a proper subvariety.

Following Proposition \ref{prop:OSM_ident_asRLCM}, suppose that $(\mathcal{H}, \bm{\pi}_{\bzeta})$ is strictly identifiable on $\mathcal{T} ~\backslash ~ \mathcal{V}$. We claim that $(\mathcal{A},\mathcal{S})$ is strictly identifiable on $\mathcal{P}$, defined as the complement of the pre-image of $\mathcal{V}$ in the parameter space of $(\mathcal{A},\mathcal{S})$. Suppose there exists $(\mathcal{A},\mathcal{S}) \neq (\bar{\mathcal A},\bar{\mathcal S}) $ on $\mathcal{P}$ such that 
\begin{align*}
  P_{\text{step}}(\bY = \yy ;\, \mathcal A, \mathcal S , \bm{\Sigma})= P_{\text{step}}(\bY = \yy  ;\, \bar{\mathcal A}, \bar{\mathcal S} , \bm{\Sigma}), \!\quad \text{ for all } \yy \in \{0,1\}^{J}.
\end{align*}

By \eqref{eq:OSM_equiv_RLCM-step} and Proposition \ref{prop:OSM_ident_asRLCM}, we have $\mathcal{H} = \bar{\mathcal{H}}$, where $\mathcal{H}$ and $\bar{\mathcal{H}}$ correspond to $(\mathcal{A},\mathcal{S})$ and $(\bar{\mathcal A}, \bar{\mathcal S})$, respectively. In particular, we have 
\begin{align}\label{eq:OSM_eta-equiv}
    \eta_{j, \bzeta} = \bar{\eta}_{j, \bzeta}, \!\quad \text{ for all }\!\quad j \ \text{and} \ \bzeta. 
\end{align}

For any $k \in \{1, \dots, K\}$ and $j \in \{1, \dots, J\}$ such that $q_{jk} =1$, consider $\bzeta$ with $\zeta_k=L+1$ and $\zeta_k^{'} = 1$ for any $k^{'} \neq k$. Since $s_{jk,L+1}=1$ and $s_{jk^{'},1}=0$, it follows from \eqref{eq:OSM_etajzeta} and \eqref{eq:OSM_eta-equiv} that $\alpha_{jk} = \bar{\alpha}_{jk}$. For $l\in \{2, \dots, L\}$, consider $\bzeta$ such that $\zeta_k=l$ and $\zeta_{k'} = 1$ for any $k^{'} \neq k$. Then we have $\alpha_{jk} s_{jk,l} = \bar{\alpha}_{jk}  \bar{s}_{jk,l},$
which implies $s_{jk,l} = \bar{s}_{jk,l}$. Since $k$ and $j$ are arbitrary, we have $(\mathcal A,\mathcal S) = (\bar{\mathcal A}, \bar{\mathcal S})$. We now proceed to prove Proposition \ref{prop:OSM_ident_asRLCM}.
\begin{proof}[Proof for Proposition \ref{prop:OSM_ident_asRLCM}]
    The proof follows from the argument in the proof of Theorem 4.3 presented in \cite{Gu&Xu_AoS2020}, which considers an RLCM where each coordinate $\zeta_k$ takes only two possible values. Most of the arguments extend naturally to our setting. Therefore, we present only the key components that require adaptation. 
    
    Define the marginal probability matrix $T(\mathcal{H})$ of size $2^{J} \times(L+1)^{K}$, where $J$ denotes the number of items and $(L+1)^{K}$ denotes the number of classes. Rows of $T(\mathcal{H})$ are indexed by the $2^J$ possible response patterns $\yy \in \{0,1\}^{J}$ and columns of $T(\mathcal{H})$ are indexed by the latent classes $\bzeta  \in \{1, \dots, L+1\}^{K}$. The $(\yy, \bzeta)$-th entry of $T(\mathcal{H})$, denoted by $T_{\yy, \bzeta}(\mathcal{H})$, represents the marginal probability that subjects in latent class $\bzeta$ provide positive responses to the set of items $\{j : y_j = 1\}$, namely
\begin{align*}
    T_{\yy, \bzeta}(\mathcal{H}) = P(\bY \succeq \yy ;\, \mathcal{H}, \bzeta) = \prod_{j: y_j = 1} \eta_{j, \bzeta},
\end{align*}
where, for $\boldsymbol{a}, \boldsymbol{b} \in \mathbb{R}^{d}$, we write $\boldsymbol{a} \succeq \boldsymbol{b}$ if $a_i \geq b_i$ for all $i =1, \dots, d.$

Without loss of generality, assume that the items have been arranged according to Condition 1 such that $\mathbf{Q}^{\top} = (\mathbf{Q}_1^{\top}, \mathbf{Q}_2^{\top},(\mathbf{Q}')^{\top})$. With a slight abuse of notation, for $i\in \{1,2\}$, let $T(\mathbf{Q}_i, \mathcal{H}_{\mathbf{Q}_i})$ denote the $2^{LK} \times (L+1)^{K}$ $T$-matrix. Following the argument in \cite{Gu&Xu_AoS2020}, to show generic identifiability, it suffices to find one specific set of item parameters $\mathcal{H}$ satisfying the constraints imposed by the \textbf{Q}-matrix that make the $T$-matrices $T(\mathbf{Q}_i, \mathcal{H}_{\mathbf{Q}_i})$ full rank. In the following we focus on $T(\mathbf{Q}_1, \mathcal{H}_{\mathbf{Q}_1})$ only. 

Recall that for $\mathbf{Q}_1 = (q_{1,jk})_{(LK) \times K}$, C1 implies that $q_{1,(c-1)K +k,k} = 1$ for $k = 1, \dots, K,\, c = 1, \dots, L$. We specify the parameters $s_{jk,l}$ such that $s_{jk,l}=0$ if $j-k$ is not a multiple of $K$. By \eqref{eq:OSM_etajzeta}, we have
\begin{align*}
    \eta_{(c-1)K +k, \bzeta} =  s_{(c-1)K +k,k,\zeta_k}, \!\quad \text{for}\!\quad c = 1, \dots L;\   k = 1, \dots, K. 
\end{align*}
We further define $s_{(c-1)K +k,k,l} = \bm 1(l \geq L+2-c)$, for $c = 1, \dots L;\   k = 1, \dots, K;\ l = 1, \dots, L$. For each latent class $\bzeta$, define $S(\bzeta):=\{j:j \in \{1, \dots, LK\};\, \eta_{j,\bzeta}=1\}.$ Under this construction, it can be verified that the sets $S(\bzeta)$ are all distinct. We reorder the columns of the $T$-matrix so that they are indexed by $\bzeta^{(1)}, \dots, \bzeta^{(L+1)^{K}}$, with the property that $S(\bzeta^{(i)}) \not\subseteq  S(\bzeta^{(j)})$ for $i < j$. For each $i$, define $\yy^{(i)} = (y_1^{(i)}, \dots, y_{LK}^{(i)})\in \{0, 1\}^{LK}$ such that 
\begin{equation*}
y_j^{(i)}=
\begin{cases}
1,& j\in S(\bzeta^{(i)}),\\
0,& j\notin S(\bzeta^{(i)}).
\end{cases}    
\end{equation*}

Now consider the square sub-matrix $M$ whose rows are indexed by $\yy^{(1)},\dots,\yy^{((L+1)^K)}$ and columns by $\bzeta^{(1)},\dots,\bzeta^{((L+1)^K)}$. Its $(i,j)$-th entry is
\begin{equation*}
M_{ij} = T_{\yy^{(i)},\bzeta^{(j)}} = \bm 1(S(\bzeta^{(i)})\, \subseteq\, S(\bzeta^{(j)}) )
\end{equation*}

Hence, $M_{ii}=1$ for all $i$, and $M_{ij}=0$ if $j>i$. Therefore, $M$ is lower triangular with diagonal entries equal to 1, and hence invertible. It follows that $T(\mathbf{Q}_1,\mathcal H_{\mathbf{Q}_1})$ has full column rank. 

The remainder of the proof follows from the argument in Theorem 4.3 of \cite{Gu&Xu_AoS2020}, which in turn relies on the uniqueness of three-way tensor decompositions established in \cite{Kruskal_LinAlgApp1977} and \cite{Rhodes_LinAlgApp2010}.
\end{proof}

\subsection{On the approximation of \texorpdfstring{$g_{jk}^{\text{step}}$}{} to \texorpdfstring{$g_{jk}^s$}{}}\label{sec:OSM_approx}

We introduce the following lemma
\begin{lemma} \label{prop:OSM_lemma}
Any piecewise linear, monotone non-decreasing, continuous function $g^s_{jk}$ can be approximated arbitrarily well by a piecewise constant function $g_{jk}^{\text{step}}$ on a sufficiently fine grid. 
\end{lemma}
\begin{proof}[Proof of Lemma \ref{prop:OSM_lemma}]
Since $g^s_{jk}$ is continuous on the compact interval $[0,1]$, it is uniformly continuous: for any $\epsilon > 0$, there exists $\delta > 0$ such that $|x-x'|< \delta$ implies $|g^s_{jk}(x) - g^s_{jk}(x')| < \epsilon$. Take a regular grid $0 = \kappa_0 < \kappa_1 <\cdots < 1 = \kappa_{L+1}$ with $|\kappa_{l}-\kappa_{l-1}| < \delta$ for all $l = 1,\ldots,L(\delta)$. Define
\begin{equation*}
    g_{jk}^{\text{step}}(x) = \sum_{l=1}^{L(\delta)+1} g^s_{jk}(\kappa_{l-1}) \times \bm 1(x \in [\kappa_{l-1}, \kappa_l)),\,
\end{equation*}
which is \eqref{eq:OSM_step} when $s_{jk,l} = g^s_{jk}(\kappa_{l-1})$, and thus $\text{sup}_x\, |g_{jk}^s(x)-g_{jk}^{\text{step}}(x)| < \epsilon$.
\end{proof}

Lemma \ref{prop:OSM_lemma} can be easily extended to other bases that produce monotone non-decreasing continuous functions (e.g., I-Splines \citealp{Ramsay_StatSci1988}).

\bibliographystyle{apalike}
\bibliography{bib_GaPM-CDM}